\documentclass[oneside]{elsarticle}

\usepackage{geometry}
\usepackage[utf8]{inputenc}
\usepackage{enumitem,amssymb}
\usepackage{graphicx}
\usepackage{dcolumn}
\usepackage{bm}
\usepackage{subfigure}
\usepackage{float}
\usepackage{etoolbox}
\patchcmd{\MaketitleBox}{\footnotesize\itshape\elsaddress\par\vskip36pt}{\footnotesize\itshape\elsaddress\par\parbox[b][36pt]{\linewidth}{\vfill\hfill\textnormal{\large \today}\hfill\null\vfill}}{}{}%
\patchcmd{\pprintMaketitle}{\footnotesize\itshape\elsaddress\par\vskip36pt}{\footnotesize\itshape\elsaddress\par\parbox[b][36pt]{\linewidth}{\vfill\hfill\textnormal{\large \today}\hfill\null\vfill}}{}{}%

\makeatletter
\def\ps@pprintTitle{%
  \let\@oddhead\@empty
  \let\@evenhead\@empty
  \let\@oddfoot\@empty
  \let\@evenfoot\@oddfoot
}
\makeatother

\usepackage[T1]{fontenc}
\usepackage[utf8]{inputenc}
\usepackage{xcolor}
\usepackage{amsmath}

\usepackage{tikz}
\usetikzlibrary{shapes.geometric, arrows}
\tikzstyle{startstop} = [rectangle, rounded corners, minimum width=3cm, minimum height=1cm,text centered, draw=black, fill=red!30]
\tikzstyle{io} = [trapezium, trapezium left angle=70, trapezium right angle=110, minimum width=3cm, minimum height=1cm, text centered, draw=black, fill=blue!30]
\tikzstyle{process} = [rectangle, minimum width=3cm, minimum height=1cm, text centered, draw=black, fill=orange!30]
\tikzstyle{decision} = [diamond, minimum width=3cm, minimum height=1cm, text centered, draw=black, fill=green!30]
\tikzstyle{arrow} = [thick,->,>=stealth]
\newlist{thematic}{itemize}{8}
\setlist[thematic]{label=$\square$}
\usepackage{pifont}

\title{Direct 0D-3D coupling of a lattice Boltzmann methodology for fluid-structure hemodynamics simulations}

\author[USC]{Heng Wei}
\ead{famlani@gmail.com}
\author[USC,UPS]{Faisal Amlani}
\ead{famlani@gmail.com}
\author[USC,USC2]{Niema M Pahlevan\corref{cor1}}
\ead{pahlevan@usc.edu}

\cortext[cor1]{Corresponding author}

\address[USC]{Department of Aerospace and Mechanical Engineering, University of Southern California, Los Angeles, USA}
\address[UPS]{Universit\'{e} Paris-Saclay, CentraleSup\'{e}lec, ENS Paris-Saclay, CNRS, LMPS - Laboratoire de M\'{e}canique Paris-Saclay, 91190, Gif-sur-Yvette, France.}
\address[USC2]{School of Medicine, University of Southern California, Los Angeles, USA}
\date{Draft updated: \today}

\usepackage[export]{adjustbox}% http://ctan.org/pkg/adjustbox
\definecolor{red}{rgb}{1,0,0.0}

\usepackage{algorithm}
\usepackage{algcompatible}

\renewcommand{\COMMENT}[2][.3\linewidth]{%
  \leavevmode\hfill\makebox[#1][l]{//~#2}}
\algnewcommand\algorithmicto{\textbf{to}}
\algnewcommand\RETURN{\State \textbf{return} }
\newcommand{\pd}[2]{\frac{\partial #1}{\partial #2}}
\newcommand{\bs}{\boldsymbol{s}}
\newcommand{\bx}{\boldsymbol{x}}
\newcommand{\bX}{\boldsymbol{X}}
\newcommand{\bB}{\boldsymbol{B}}
\newcommand{\bb}{\boldsymbol{b}}
\newcommand{\bv}{\boldsymbol{v}}

\usepackage{lineno}
\usepackage{setspace}
\begin{document}
\begin{abstract}{This work introduces a numerical approach and implementation for the direct coupling of arbitrary complex ordinary differential equation- (ODE-)governed zero-dimensional (0D) boundary conditions to three-dimensional (3D) lattice Boltzmann-based fluid-structure systems for hemodynamics studies. In particular, a most complex configuration is treated by considering a dynamic left ventricle- (LV-)elastance heart model which is governed by (and applied as) a nonlinear, non-stationary hybrid ODE-Dirichlet system. Other ODE-based boundary conditions, such as lumped parameter Windkessel models for truncated vasculature, are also considered. Performance studies of the complete 0D-3D solver, including its treatment of the lattice Boltzmann fluid equations and elastodynamics equations as well as their interactions, is conducted through a variety of benchmark and convergence studies that demonstrate the ability of the coupled 0D-3D methodology in generating physiological pressure and flow waveforms---ultimately enabling the exploration of various physical and physiological parameters for hemodynamics studies of the coupled LV-arterial system. The methods proposed in this paper can be easily applied to other ODE-based boundary conditions as well as to other fluid problems that are modeled by 3D lattice Boltzmann equations and that require direct coupling of dynamic 0D boundary conditions.}\end{abstract}

% \keywords{3D blood flow,  0D-3D coupling, lattice Boltzmann equations, fluid-structure interactions, mathematical physiology, LV-arterial hemodynamics}

\maketitle

\section{{Introduction}}\label{sec:introduction}

Cardiovascular modeling is a challenging fluid-structure interaction problem that involves treatment of complex geometries and boundary conditions in order to effectively capture physiological dynamics~\cite{taylor2009patient}. Computational fluid dynamics (CFD) is a widely-used approach for simulating blood flow in the circulatory system~\cite{taylor2010image,pennati2010modeling,vignon2010primer,mittal2001application,iman1,iman2,seo2017method,wu2015coupled,amlanipahlevan,aghilinejad2020dynamic}, which includes applications to 1D~\cite{amlanipahlevan,aghilinejad2020dynamic,shi2011review}, 2D~\cite{tasciyan1993two} or 3D~\cite{mittal2001application,iman1,iman2,seo2017method,wu2015coupled,steinman2005flow,steinman2002image} formulations. The lattice Boltzmann (LB) method~\cite{timm2016lattice, benzi1992lattice,shan1993lattice,fang2002lattice}, originating from classical statistical physics, is a powerful alternative to conventional continuum-based CFD methods that use Navier-Stokes equations. The LB method uses simplified kinetic equations combined with a modified molecular-dynamics approach to model both Newtonian and non-Newtonian fluid flow in any complex geometry (the fluid is modeled as particles that stream and collide over a discrete lattice mesh). Indeed, a particular advantage of LB-based hemodynamics solvers is their ability to easily model non-Newtonian effects via its right-hand-side; capturing such effects may be important for small vessels or vessels where the shear rate is low~\cite{sriram2014non,sankar2007non}. The accuracy and usefulness of the LB method have been demonstrated in a variety of fluid dynamics problems including turbulence~\cite{cosgrove2003application} and multiphase flow~\cite{shan1993lattice}. As highlighted in previous studies~\cite{shan1993lattice,fang2002lattice,cosgrove2003application,boyd2007analysis,wei2020significance}, LB methods have been shown to be particularly suitable for hemodynamics simulations since many flow features of clinical interest may require efficient numerical treatment of fully 3D computational domains.
%
% where the shear rate \red{add here the advantage of LB for hemodynamics, e.g., non-newtonian effects easily as RHS (although not important for large vessel, however, where the shear rate is low or in smaller vessels)}

In order to extend the clinical applicability of fluid-structure blood flow solvers based on LB equations applied to large vessels, this work introduces a direct 0D-3D coupling for the treatment of physiological boundary conditions that are governed by ordinary differential equations (ODEs) such as lumped parameter Windkessel models~\cite{westerhof1969analog,vignon2004outflow} or more complex hybrid ODE-Dirichlet systems such as time-varying elastance organ models~\cite{amlanipahlevan} . Previous contributions on the 0D-3D coupling for finite element methods~\cite{moghadam2013modular,kim2009coupling} have been implicit and iterative, and for lattice Boltzmann~\cite{moore2018towards,sadeghi2020towards} blood flow models usually only a Dirichlet or Neumann pressure or flow is prescribed during the entirety of a cardiac cycle (precluding the use of more sophisticated and non-stationary, i.e., switching, boundary conditions~\cite{amlanipahlevan}).

Additionally, recent work~\cite{sadeghi2020towards} on LB-based hemodynamics solvers have assumed only rigid walls, and have applied 0D lumped parameter models externally through an iterative procedure where the heart model is evolved and precomputed entirely independently~\cite{sadeghi2020towards} (such that the resultant pressure profile is applied on a 3D LB domain simply as a Dirichlet condition, i.e., not a true mathematical coupling).

% 0D-coupling for LB-based hemodynamics are not directly coupled

 This work, on the other hand, presents a first direct 0D-3D coupling for fully fluid-structure 3D pulsatile blood flow solvers based on LB and elastodynamics equations. In particular, the 0D equations considered in this work govern a highly-complex and non-stationary dynamic left ventricle (LV-)elastance heart model~\cite{amlanipahlevan} (that switches between an ODE and a Dirichlet boundary condition in a ``non-stationary'' fashion~\cite{amlanipahlevan}) in order to generate physiologically-accurate hemodynamic conditions (instead of simply assigning a given inlet flow or pressure, as is commonly done~\cite{sadeghi2020towards,vignon2006outflow,vignon2006outflownon}). Such a coupled model represents the most complicated boundary configuration found in the circulatory system~\cite{amlanipahlevan}: a hybrid ODE-Dirichlet boundary condition representing the left ventricle, where the time at which the ODE-governed condition transitions to a Dirichlet condition is itself determined by the corresponding solution of the governing fluid-structure LB system. Hence the methodology introduced in this work can be trivially extended to the application of non-switching ODE-based boundary conditions such as lumped parameter models based on Windkessels~\cite{westerhof1969analog,vignon2004outflow,amlanipahlevan} (also treated in this contribution).

This work presents a numerical approach for directly coupling these 0D LV-elastance and Windkessel boundary conditions (or any simpler ODE-based boundary condition) to a 3D LB-based fluid-structure interaction solver for hemodynamics (where the solid is governed by elastic equations). {The methodology introduces, for both ODE-based as well as non-stationary boundaries, a discrete explicit-in-time extension to an LB non-equilibrium extrapolation method~\cite{zhao2002non,timm2016lattice,kang2010non} that has been previously proposed for fluid-only problems (i.e., no solid interaction) and only for given (often analytical) Dirichlet-based pressure/velocity boundary conditions.} The ultimate aim is to enable accurate physiological LV-aortic coupling conditions for cardiovascular studies of, for example, the effects of left ventricle contractility on pulsatile hemodynamics in the aorta. Section~\ref{sec:section2} presents the governing equations for the fluid, the solid and the LV-elastance models. Section~\ref{sec:coupling} details the direct 0D-3D LV-coupling strategy that is introduced in this work, including a discussion of the loss of mathematical regularity of such a model from a discontinuity in the velocity upon valve closure (and the proposition of a smoothing operator in order to ensure a continuous transition). Section~\ref{sec:section4} provides algorithmic details of the complete solver, including the numerical methods employed for the solid as well as the fluid-structure interactions (both of which can be provided by any number of suitable schemes). Section~\ref{sec:performance} presents a variety of performance studies attesting to the valid implementation and the accuracy of the fluid and solid solvers presented in this paper. Finally, Section~\ref{sec:casestudy} considers a sample physiological study of oscillatory wall shear stress in a 3D aorta with carotid and renal branches.

\section{Materials and Methods}\label{sec:methods}

\subsection{{Governing formulations}}\label{sec:section2}

\begin{figure}
\centering
\includegraphics[width=0.65\textwidth]{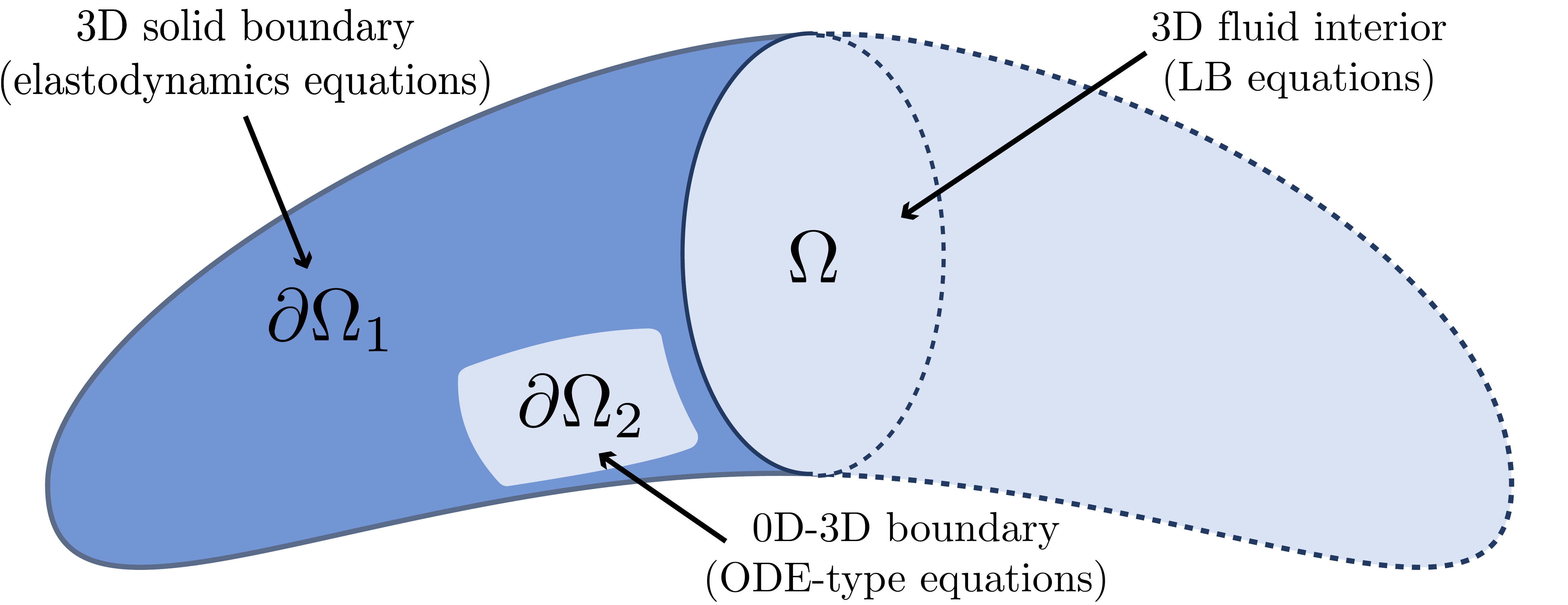}% Here is how to import EPS art
\caption{\label{fig:domain} A representative illustration of the complete 3D computational domain defined by $\overline{\Omega} = \Omega \cup \partial \Omega_1 \cup \partial \Omega_2$, where $\Omega$ denotes the fluid interior (governed by lattice Boltzmann equations), $\partial \Omega_1$ denotes the compliant solid boundary (governed by elastodynamics PDEs and incorporated by any appropriate fluid-structure interaction algorithm), and $\partial \Omega_2$ denotes the coupled 0D-3D boundary (governed by time-dependent ODEs).}
\end{figure}

This section presents the governing equations employed in the numerical solver described in Section~\ref{sec:section4}: those for the fluid domain of a vessel (governed by lattice Boltzmann equations, Section~\ref{sec:governingfluid}); those for the LV-elastance model for the fluid inlet (governed by hybrid ODE-Dirichlet equations, Section~\ref{sec:governingLV}); and those for the solid vessel walls (governed by elastodynamics equations, Section~\ref{sec:governingsolid}). An illustration of the complete coupled fluid-structure computational domain $\overline{\Omega} = \Omega \cup \partial \Omega_1 \cup \partial \Omega_2\in\mathbb{R}^3$ is presented in Figure~\ref{fig:domain} for the (interior) fluid domain $\Omega$, the solid wall $\partial \Omega_1$ and the 0D-3D coupled domain $\partial \Omega_2$.

\subsubsection{3D lattice Boltzmann equations\label{sec:governingfluid}}

For the fluid domain $\Omega$, the lattice Boltzmann (LB) equations are employed, where the synchronous motions of fluid particles on a regular lattice are enforced through a particle distribution function~\cite{timm2016lattice}. This distribution function enforces mass and momentum conservation as well as ensuring that the fluid is Galilean invariant and isotropic \cite{wolf2004lattice}. In the present work, a single-relaxation-time (SRT) incompressible LB method is used to solve the incompressible flow \cite{he1997lattice}. The evolution of the distribution functions on the lattice is governed by the discrete Boltzmann equation with the Bhatnagar-Gross-Krook (BGK) collision model, given by
\begin{eqnarray}
\label{Eqn:fi}
f_i(\boldsymbol{x}+\boldsymbol{e}_i\Delta t,t+\Delta t)-f_i(\boldsymbol{x},t)=-\frac{1}{\tau}[f_i(\boldsymbol{x},t)
-f_i^{eq}(\boldsymbol{x},t)]+\Delta t F_i(\bx,t), \quad i=0,...,N_0-1,
\end{eqnarray}
where $f_i(\bm{x},t)$ are distribution functions of the particles in phase space; $\bm{e}_i$ are discrete velocities at position $\bm{x}$ and time $t$; $\tau$ is a non-dimensional relaxation time; $f_i^{eq}(\bx,t)$ are equilibrium distribution functions; and $F_i$ are forcing terms. Here, $N_0=19$ since a D3Q19 (19 discrete velocity vectors) stencil is applied (and a D2Q9 stencil is employed for the 3D-axisymmetric cases, i.e., $N_0=9$.). The non-dimensional relaxation time $\tau$ is related to fluid viscosity $\mu$ by the expression
\begin{eqnarray}
\label{Eqn:tau}
\mu=\rho \nu=\rho c_s^2\left(\tau-\frac{1}{2}\right)\Delta t,
\end{eqnarray}
where $\nu$ is the kinematic viscosity, $\rho$ is the incompressible fluid density (e.g., blood density), and $c_s={\Delta x}/({\Delta t\sqrt{3}})$ is the lattice sound speed. Uniform discretizations are employed throughout this work for both time ($\Delta t$) and lattice space ($\Delta x$), chosen such that ${\Delta x}/{\Delta t}=1$ (corresponding to $c_s=1/\sqrt{3}$).

The equilibrium distribution functions $f_i^{eq}(\bm{x},t)$ for an incompressible Lattice Boltzmann model~\cite{he1997lattice} and the forcing terms  $F_i(\bx,t)$~\cite{he1998discrete} are resepctively defined as
\begin{eqnarray}
\label{Eqn:feq}
f_i^{eq}(\bx,t)= \frac{\omega_i P(\bx,t)}{c_s^2}+\omega_i\rho\left[\frac{\boldsymbol{e}_i\cdot\boldsymbol{v}(\bx,t)}{c_s^2}+
\frac{(\boldsymbol{e}_i\cdot\boldsymbol{v}(\bx,t))^2}{2c_s^4}-\frac{\boldsymbol{v}(\bx,t)^2}{2c_s^2} \right] \quad \text{and}
\end{eqnarray}
\begin{eqnarray}
\label{Eqn:f}
F_i(\bx,t)=\left(1-\frac{1}{2\tau}\right)\times\omega_i\times\left[\left(\frac{{\boldsymbol{e}_i-\boldsymbol{v}(\bx,t)}}{c_s^2}+\frac{\boldsymbol{e}_i\cdot\boldsymbol{v}(\bx,t)}{c_s^4}\boldsymbol{e}_i\right)\cdot\boldsymbol{b}(\bx,t)\right],
\end{eqnarray}
where $\bv(\bx,t)$ is the fluid velocity; ${P}(\bx,t)$ is the macroscopic pressure; $\omega_i$ are weighting factors (whose values are adopted from previous study~\cite{he1997lattice}); and $\bm{b}(\bm{x},t)$ is the force density in Eulerian coordinates. Pressure ${P}(\bx,t)$ and velocity $\bm{v}(\bx,t)$ can be calculated from the distribution functions $f_i(\bx,t)$ via the expressions
\begin{eqnarray}
\label{Eqn:rho}
P(\bx,t)=c_s^2\sum_if_i(\bx,t),
\end{eqnarray}
\begin{eqnarray}
\label{Eqn:v}
\boldsymbol{v}(\bx,t)=\frac{1}{\rho}\sum_i\boldsymbol{e}_if_i(\bx,t)+\frac{1}{2\rho}\boldsymbol{b}(\bx,t)\Delta t.
\end{eqnarray}
The  numerical implementations and algorithmic pseudocode are provided in Section~\ref{sec:section4}.
\subsubsection{0D fluid boundary equations\label{sec:governingLV}: LV-elastance and lumped parameter models}

% \subsubsection{A non-stationary LV-elastance heart model}

For the fluid inlet boundary condition on the coupled 0D-3D interface $\partial \Omega_\text{inlet} \subset \partial \Omega_2$, when the heart valve is open, the corresponding time-dependent boundary condition for $\bx\in\Omega_\text{inlet}$ that governs the dynamics of the left ventricle is given by the ODE~\cite{berger,amlanipahlevan}
\begin{eqnarray}
\label{Eqn:LVE}
\frac{\partial {P_v}}{\partial t}=-\frac{1}{C_v(t)}
\Bigg[\frac{\partial C_v(t)}{\partial t}P_v(t)+Q(\bx,t)\Bigg],
\end{eqnarray}
for pressure inside the ventricle $P_v(t)$ and time-varying compliance $C_v(t)$ (the inverse of which is the corresponding time-varying elastance whose maximum is a measure of LV contractility~\cite{berger,amlanipahlevan}). Hence once the pressure $P_v(t)$ in the ventricle is greater than that of the inlet fluid domain boundary $P(\bx\in\partial\Omega_\text{inlet}, t)$, the valve opens and $P(\bx, t)=P_v(t)$ with the corresponding flow condition $Q(t)$ given naturally by the fluid solver via the area integral given by
\begin{equation}
Q(t) = \iint_{\partial\Omega_2} \frac{\sum \bm{e}_i f_i(\bx,t)}{\sum f_i(\bx,t)} dA.
\end{equation}
Once the inflow reaches zero (or, numerically, the time at which $Q(t) < 0$), the valve closes and the 0D boundary condition remains $Q(t)=0$ (a Dirichlet-type flow condition). Generally, $C_v(t)$ is given by clinical parameters either through a look-up table or through a closed-form approximation (adopted throughout this paper from the compliance curve presented in Amlani and Pahlevan~\cite{amlanipahlevan}).  An analysis of this \emph{non-linear}, \emph{non-stationary} LV boundary condition and details on the algorithmic implementation of such a switching configuration are provided in Section~\ref{sec:coupling}.

% \subsubsection{A lumped-parameter model for eliminated periphery}

At the outlet boundary $\partial \Omega_\text{outlet} \subset \partial \Omega_2$ of the 3D physiological aorta considered in Section~\ref{sec:casestudy}, a conventional 0D lumped parameter model, a so-called circuit-like \emph{Windkessel} model, is employed to represent the effects of truncated vasculature~\cite{amlanipahlevan} (i.e., eliminated periphery). The outlet of the fluid domain is coupled to such a model through a matching characteristic impedance $Z_w$ that is related to the fluid inductance and overall outgoing aortic compliance. Together with an effective chamber compliance $C_w$ and a total peripheral resistance $R_w$, the pressure $P_w$ in the terminal compliance chamber is related to the aortic pressure $P(\bx,t)$ at the outlet boundary $\bx\in\partial\Omega_\text{outlet}$ through an ODE given by
\begin{equation}\label{eq:rightbc}
  \pd{P_w}{t}(t) = \frac{1}{C_w Z_w}P(\bx,t) - \frac{R_w+Z_w}{C_wR_wZ_w} P_w(t).
\end{equation}
 The corresponding 0D outflow $Q(\bx,t)$ at the outlet $\bx\in\partial\Omega_\text{outlet}$ is given by
\begin{equation}\label{eq:rightQ}
  Q(\bx,t) = \frac{1}{Z_w} \left(P(\bx,t) - P_w(t)\right).
\end{equation}
Further details on the parameters, usage and implementation of both this outgoing Windkessel model, as well as the LV-elastance heart model above, can be found elsewhere~\cite{amlanipahlevan}.

\subsubsection{3D elastic wall equations\label{sec:governingsolid}}

In order to account for fluid-structure interactions in a vessel, the solid boundary $\partial\Omega_1$ is assumed to be a thin wall that can be described by deformation of a compliant (elastic) wall in a Lagrangian coordinate system~\cite{hua2014dynamics}, i.e.,
\begin{eqnarray}
\label{Eqn:structure}
\rho_s h \frac{\partial ^2\bm{X}}{\partial t^2}=\sum_{i,j=1}^{2}\frac{\partial}{\partial s_i}
\Bigg[Eh\varphi_{ij}\left(\sqrt{\frac{\partial \bm{X}}{\partial s_i}\cdot\frac{\partial \bm{X}}{\partial s_j}}
-\delta_{ij}\right)\frac{\partial \bm{X}}{\partial s_j}-\frac{\partial}{\partial s_j}\left(EI\gamma_{ij}\frac{\partial ^2\bm{X}}{\partial s_i \partial s_j}\right)\Bigg]+\bm{B}(\boldsymbol{s},t),
\end{eqnarray}
where $\rho_{s}$ is the density of the solid wall; $h$ is a constant wall thickness; $\delta_{ij}$ is the Kronecker delta; $\bm{X}(\boldsymbol{s},t)\in\mathbb{R}^3$ is the position of the solid wall; $\boldsymbol{s}=(s_1,s_2)\in\mathbb{R}^2$ are the Lagrangian coordinates along the solid wall; $\bm{B}(\boldsymbol{s},t)$ is the Lagrangian force exerted on the solid wall by the fluid; and $Eh$, $EI$ are stretching and bending stiffnesses (respectively). The matrices $(\varphi)_{ij}$ and $(\gamma)_{ij}$ represent in- and out-of-plane effects and, for a Poisson's ratio $\hat{\nu}$, are respectively given by
\begin{equation}
  (\varphi)_{ij} = \begin{pmatrix} 1 & \dfrac{1}{2(1+\hat{\nu})}\\ \dfrac{1}{2(1+\hat{\nu})} & 1\end{pmatrix} \quad \text{and} \quad
  (\gamma)_{ij} = \begin{pmatrix} 1 & 1\\ 1 &1\end{pmatrix}.
\end{equation}
The boundary condition of the solid wall (as a shell) at a simply-supported fixed end is given by
\begin{equation}
\label{Eqn:boundaryconditionfixed}
\bm{X}=\bm{X}_0,\qquad \frac{\partial^2\bm{X}}{\partial s_i^2}=(0,0,0)^{T}, \quad i=1,2,
\end{equation}
where $\bm{X}_0$ denotes the displacement coordinates of the fixed boundaries.

\subsubsection{Non-dimensionalization}
All the above formulations for both the fluid and the solid can be non-dimensionalized via the reference quantities $\rho$, $U_\infty=Q/A$ ($A$ is the inlet area of $\partial \Omega_2$) and effective length $L$ (which is equal to a diameter $D$ if the inlet is cylindrical). The corresponding non-dimensional parameters considered in our simulations are given by Reynolds number $\text{Re}=\rho U_\infty L/\mu$, Womersley number $\text{W}_o=\sqrt{\rho \omega L^2/4\mu}$,
bending coefficient $K=EI/\rho U_\infty^2L^3$, tension coefficient $S=Eh/\rho U_\infty^2L$, and mass ratio of the solid wall to the fluid given by $M=\rho_s h/\rho L$.

\subsection{A direct 0D-3D coupling for ODE-based boundary equations and lattice Boltzmann solvers}\label{sec:coupling}
This section proposes a discrete, direct methodology for the coupling of 3D lattice Boltzmann equations with dynamic (ODE-based) 0D models which, as described before, are often found in inflow and outflow conditions for cardiovascular configurations. The hybrid ODE-Dirichlet system governed by {Equation~\eqref{Eqn:LVE}} represents a most complex form of the myriad such formulations found in cardiovascular modeling, and its particular treatment is discussed in Section~\ref{sec:particularcoupling} (although the method proposed in what follows is straightforwardly applicable to other ODE-based 0D boundary equations such as Windkessel models). The general strategy for such multidimensional coupling of the solver presented in this work is based on the \emph{non-equilibrium extrapolation method}~\cite{timm2016lattice,zhao2002non}, which has been introduced as an alternative to typical ``bounce-back'' methods~\cite{cornubert1991knudsen} used to implement (given) pressure and velocity boundary conditions for LB methods in order to preserve a consistent order-of-accuracy between the boundaries and the order-of-accuracy inherent to LB formulations~\cite{zhao2002non}. In particular, such a method is ideally suited for curved fluid boundaries~\cite{kang2010non,guo2002extrapolation} (such as those provided by fluid-structure interfaces of interest in this work), where the physical boundary need not coincide with the regular fluid lattice.

In this contribution, we present a discrete explicit-in-time LB extension of the non-equilibrium extrapolation method for the use of ODE-based boundary conditions (previous usages of non-equilibrium extrapolation have been mostly confined to given Dirichlet-based pressure or velocity boundary conditions of the fluid system~\cite{timm2016lattice,zhao2002non,kang2010non}). In order to update the distribution functions from a timestep $t_n$ to a timestep $t_{n+1}=t_n+\Delta t$ of a fluid point $\bx_f\in \Omega$ that is streamed from the corresponding 0D-3D interface (boundary) node $\bm{x}\in \partial\Omega_2$ of the lattice (i.e., $\bx_f = \bx+\boldsymbol{e}_i\Delta t$), such a formulation can be derived by separating collision and streaming operations of Equation~\eqref{Eqn:fi} into
 \begin{equation}\label{eq:collision}
   f_i^*(\boldsymbol{x})=f_i(\boldsymbol{x},t_n)-\frac{1}{\tau}[f_i(\boldsymbol{x},t_n)
   -f_i^{eq}(\boldsymbol{x},t_n)]+\Delta t F_i(\bx,t_n), \quad \bm{x} \in \partial \Omega_2,
 \end{equation}
 and
\begin{equation}
  \label{eq:streamstep}
  f_i(\boldsymbol{x}_f,t_{n+1}) =   f_i(\bx+\boldsymbol{e}_i\Delta t,t+\Delta t) = f_i^*(\boldsymbol{x}), \quad \bm{x} \in \partial \Omega_2, \quad \bm{x}_f \in \Omega,
\end{equation}
respectively. The distribution function for $\bx \in \partial \Omega_2$ in Equation~\eqref{eq:collision} can be decomposed into an equilibrium and non-equilibrium part, i.e.,
\begin{equation}\label{eq:decomposition}
  f_i(\boldsymbol{x},t_n) = f_i^{eq}(\boldsymbol{x},t_n) + f_i^{neq}(\boldsymbol{x},t_n),
\end{equation}
which, upon substitution into Equation~\eqref{eq:collision} and re-ordering terms, gives the expression for the collision as
\begin{equation}\label{eq:collisionsplit}
  f^*_i(\boldsymbol{x})=f_i^{eq}(\boldsymbol{x},t_n)+\left(1-\frac{1}{\tau}\right) f_i^{neq}(\boldsymbol{x},t_n) +\Delta t F_i(\bm{x},t_n).
\end{equation}
The non-equilibrium component on the boundary can be approximated via a Chapman-Enskog asymptotic expansion~\cite{zhao2002non} at a distance $\epsilon>0$ from the neighboring fluid node $\bm{x}_f$ as
\begin{equation}\label{eq:collisionexpansion}
  f_i^{neq}(\boldsymbol{x},t_n)=f_i (\bx_f,t_n) -f_i^{eq}(\boldsymbol{x}_f,t_n) + \mathcal{O}(\epsilon^2),
\end{equation}
whose second-order accuracy in $\epsilon$ is consistent with the order of the LB method.  Inserting this expression into Equation~\eqref{eq:decomposition} gives a second-order approximation of the distribution function on the boundary node $\bx$ as
\begin{equation}\label{eq:decomposition2}
  f_i(\boldsymbol{x},t_n) = f_i^{eq}(\boldsymbol{x},t_n) + f_i (\bx_f,t_n) -f_i^{eq}(\boldsymbol{x}_f,t_n).
\end{equation}
From the definition of the equilibrium distribution function given in Equation~\eqref{Eqn:feq}, the above expression yields a ``post-stream'' state given by
\begin{eqnarray}
\label{Eqn:Non-equilibrium}
  f_i(\boldsymbol{x},t_n) = f_i^{eq}\left(P^*(\bx),\bv^*(\bx)\right) + f_i (\bx_f,t_n) -f_i^{eq}(P(\bx_f,t_n),\bv(\bx_f,t_n)),
\end{eqnarray}
where $P^*(\bx), \bv^*(\bx)$ are the (unknown) effective pressure or velocity at the boundary point (given by the boundary condition, as described shortly). Substitution into the collision governed by Equation~\eqref{eq:collision} yields the complete post-collision stream update of the boundary values as
\begin{eqnarray}
\label{Eqn:Non-equilibrium2}
  f^*_i(\boldsymbol{x}) = f_i^{eq}\left(P^*(\bx),\bv^*(\bx)\right) + \left(1-\frac{1}{\tau}\right) (f_i (\bx_f,t_n) -f_i^{eq}(P(\bx_f,t_n),\bv(\bx_f,t_n))+\Delta t F_i(\bx_f,t_n),
\end{eqnarray}
and hence, from Equation~\eqref{eq:streamstep}, gives complete update of the fluid point $\bx_f$ from the neighboring boundary value $\bx$ as
\begin{eqnarray}
\label{Eqn:Non-equilibrium3}
  f_i(\bx_f,t_{n+1}) = f_i^{eq}\left(P^*(\bx),\bv^*(\bx)\right) + \left(1-\frac{1}{\tau}\right) (f_i (\bx_f,t_n) -f_i^{eq}(P(\bx_f,t_n),\bv(\bx_f,t_n))+\Delta t F_i(\bx_f,t_n).
\end{eqnarray}
Other non-equilibrium extrapolation implementations have mostly considered purely fluid domains, and hence do not consider a forcing $F_i$ term in the derivation. However, for the complete solver of this work, such forces (although small in an elastic regime) can be nonzero close to a fluid-structure interface.

\begin{figure}[h]
\centering
\includegraphics[width=.7275\textwidth]{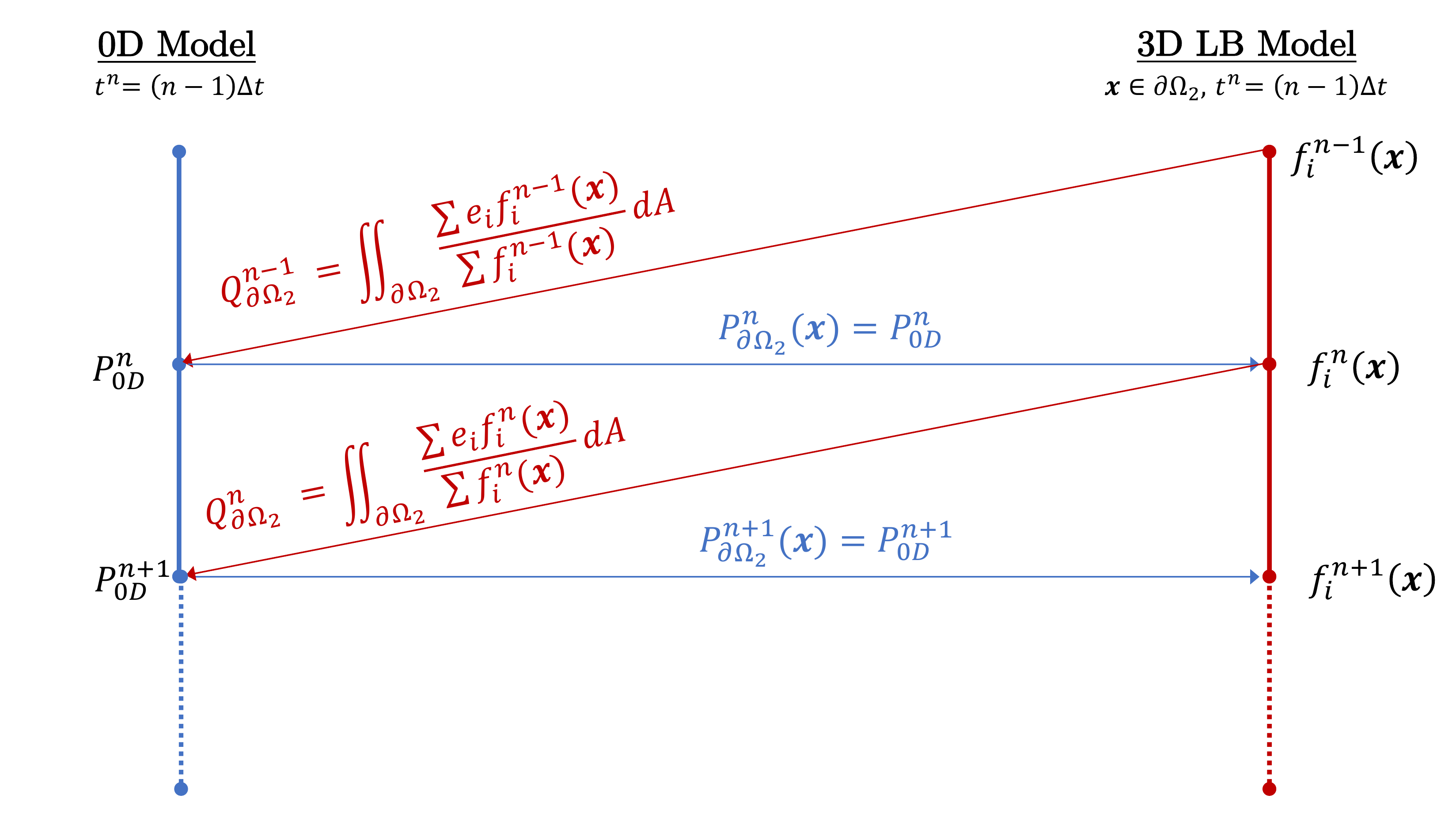}% Here is how to import EPS art
\caption{\label{fig:eout} An illustrative example diagram of the direct (explicit) coupling between a 0D model (e.g., the ODEs corresponding to the LV-heart model and the Windkessel model) and the 3D lattice Boltzmann (LB) model. The flow in $\partial \Omega_2$ is computed in terms of the LB distribution functions at a timestep $n$ which is fed into ODEs governing the 0D model. The corresponding pressure produced by the 0D model is then re-translated into distribution functions on the boundary via the non-equilibrium extrapolation described by Equation~\eqref{Eqn:Non-equilibrium}. }
\end{figure}

A 0D model (correspondingly evolved in time, if an ODE) can then be directly and explicitly coupled with this formulation at each timestep through $P^*$ and $\bv^*$. That is, if the 0D model generates a pressure $P_\text{0D}(t)$ from the flow solution (e.g., the LV-elastance model of this work), the coupling can be instituted in Equation~\eqref{Eqn:Non-equilibrium3} as
\begin{eqnarray}
(P^*(\bx), \bv^*(\bx)) = (P_\text{0D}(t_{n+1}), \bv(\bx_f,t_n)),
\end{eqnarray}
where $P^*$ comes from the evolved 0D boundary condition (e.g., an ODE), and $\bv^*$ is approximated by the corresponding value of the streamed fluid node $\bx_f$ at the current timestep. Similarly, if the 0D model generates a velocity $\bv_\text{0D}(t)$ at the boundary, then
\begin{eqnarray}
(P^*(\bx), \bv^*(\bx)) = (P(\bx_f,t_n), \bv_\text{0D}(t_{n+1})).
\end{eqnarray}
Such a direct 0D-3D coupling enables the use of any sort of mathematical or numerical model for the 0D boundary. For example, an ODE-based boundary condition can be resolved and marched forward-in-time by any suitable integration technique (e.g., forward Euler or higher-order~\cite{amlanipahlevan}). As an example 0D model that inputs a flow from the 3D fluid solver (in terms of the corresponding equilibrium function solutions) and generates a corresponding pressure (via, e.g., an ODE), an illustrative schematic diagram for such coupling as detailed above is presented in Figure~\ref{fig:eout}.

\subsubsection{On the particularities of the specific hybrid ODE-Dirichlet LV model}\label{sec:particularcoupling}

\begin{figure}
\centering
\includegraphics[width=0.485\textwidth, valign=m]{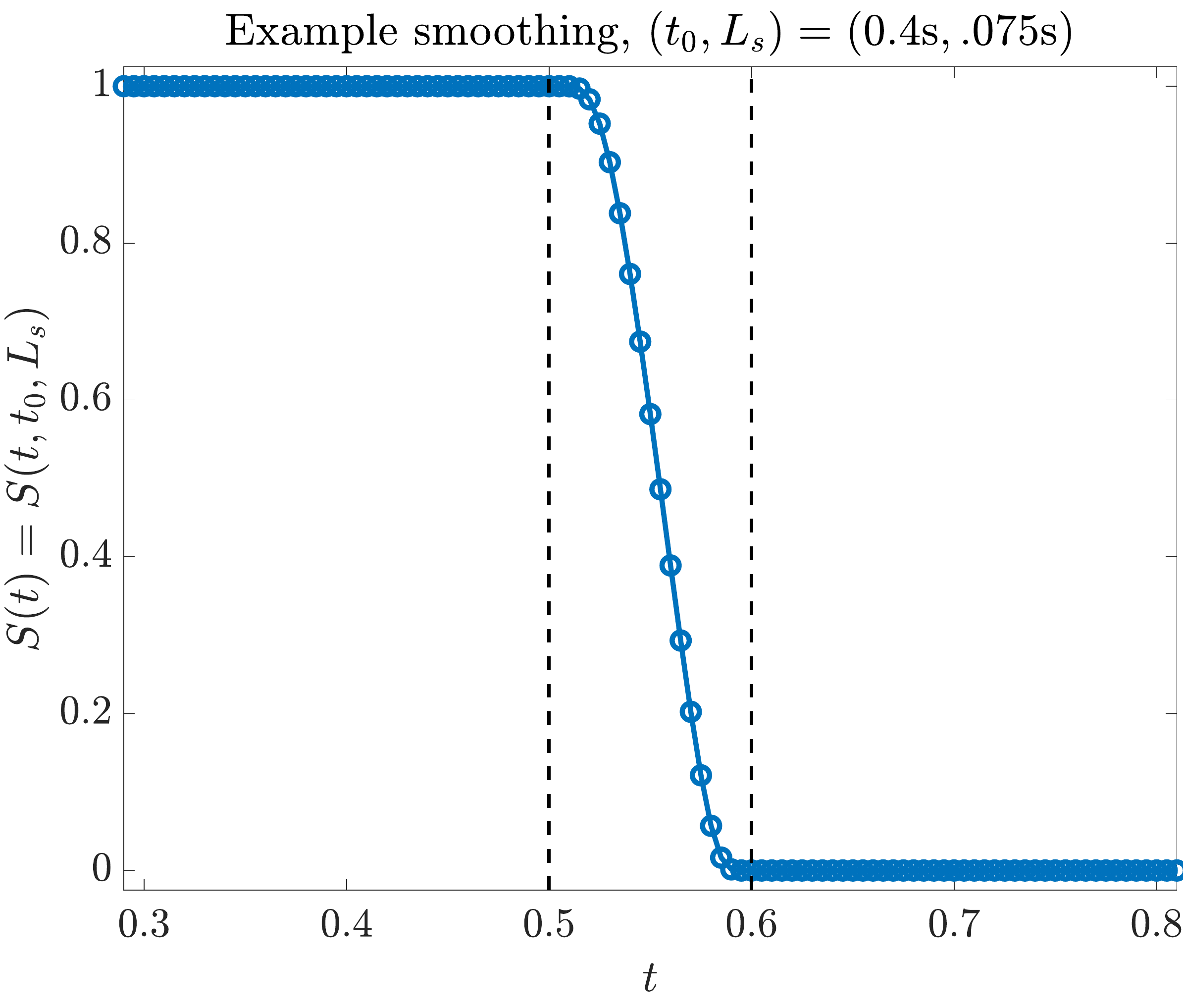}% Here is how to import EPS art
\caption{\label{fig:Smooth} An illustration of the exponential-based $C^\infty$-function $S(t,t_d,W)$ given by~\eqref{Eqn:SmoothFunction} which is employed to smoothly reduce lattice Boltzmann velocity amplitudes as the valve closes in the 0D LV-elastance model (i.e., the ODE-Dirichlet switch).}
\end{figure}

\begin{figure}
\centering
\includegraphics[width=.7275\textwidth,valign=m]{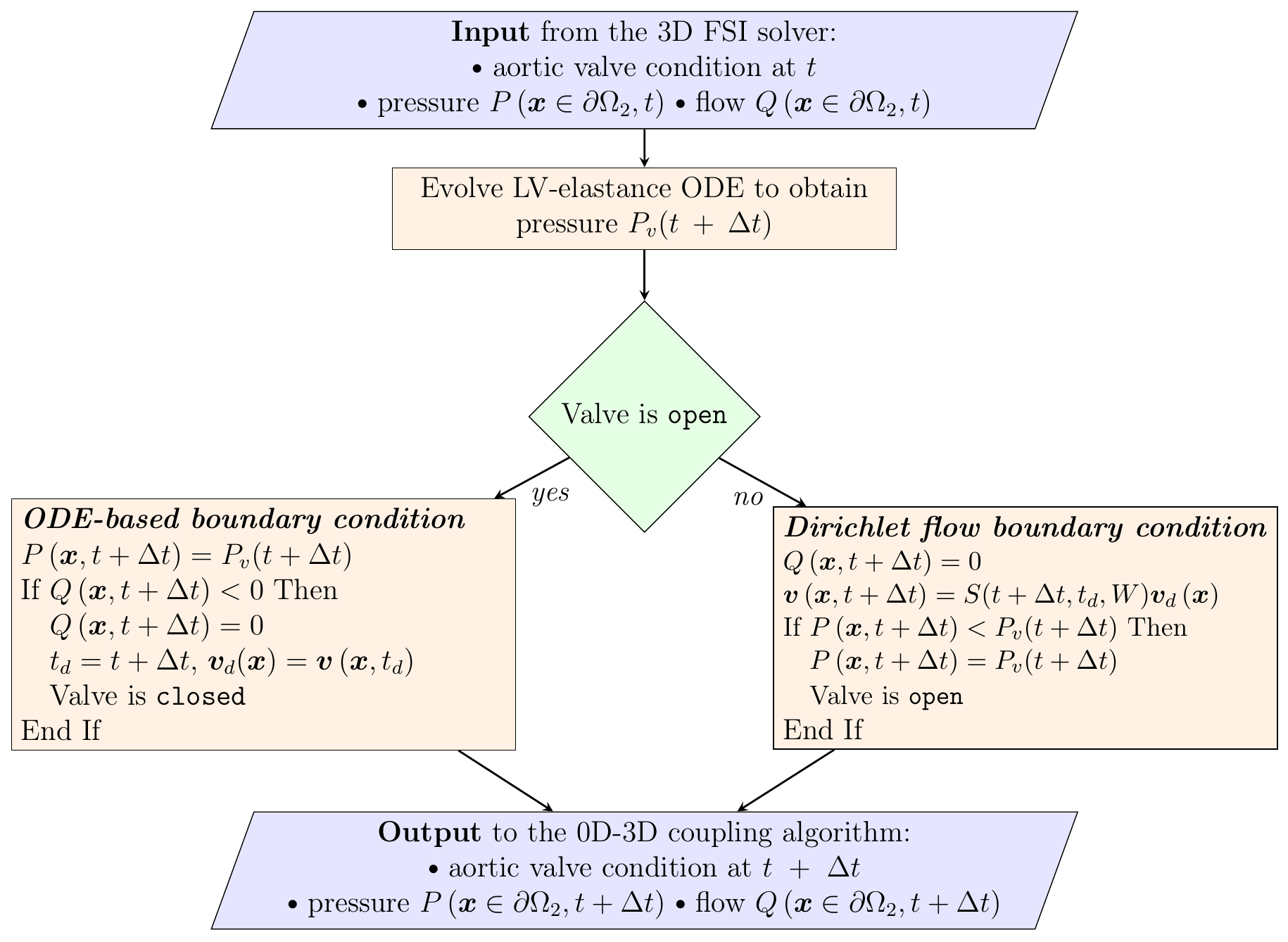}% Here is how to import EPS art
\caption{\label{fig:2} A flowchart describing the implementation of the particular hybrid ODE-Dirichlet 0D LV-elastance model that is of interest. }
\end{figure}

The non-stationary switching condition of the specific 0D LV-elastance equations presented in Section~\ref{sec:governingLV} leads to a loss in regularity of the solution near boundaries governed by the hybrid-ODE Dirichlet system. Indeed, the corresponding velocity at the time $t=T_d$ of valve closure  (also known as a dicrotic notch on the corresponding pressure waveform) is generally non-zero, and hence the switch to a Dirichlet condition for diastole leads to a discontinuity in the velocity solution close to the 0D-3D boundary $\partial \Omega_2$. That is, for a point $\bx_f\in \Omega$ neighboring a boundary node $\bx \in \partial \Omega_2$ with a velocity solution during the systolic phase ($t\in[0,T_d]$) defined as $\bm{v}_0(t) = \bm{v}(\bx_f,t) $, the velocity over a complete cardiac cycle of length $T$ can be expressed as
\begin{equation}\label{eq:voft}
  \bm{v}(t)  = \begin{cases}
  \bm{v}_{0}(t), & t \in [0,T_d] \quad (\texttt{valve open}),\\
  0, & t \in (T_d,T] \quad (\texttt{valve closed}),
\end{cases}
\end{equation}
which, again, can be discontinuous. Such a loss in the smoothness of the velocity derived from the switching solutions to the LV-elastance boundary equations may lead to spurious reflections (including in the form of artificial backflow) at the point of closure of the valve, i.e., at $t=T_d.$ In order to ensure there is no spurious backflow or artificial oscillations resulting from immediately setting a zero velocity, we introduce a smooth transition function to the velocity profile upon valve closure. Such a function can be defined as a continuously differentiable $C^\infty$ smoothing-to-zero function $S(t)$ and can be derived from an exponential or erf-based partition-of-unity as
\begin{eqnarray}
\label{Eqn:SmoothFunction}
S(t) = S(t,t_0,L_S)=
\begin{cases}
1, & t<t_0, \\
\exp\left({\dfrac{2e^{-1/u(t)}}{u(t)-1}}\right),~u(t)=\frac{t-t_0}{L_S}, & t_0\leq t\leq t_0+L_S, \\
0, & t>t_0+L_S,
\end{cases}
\end{eqnarray}
where $t_0=T_d$ is the location of the start of the smoothing-to-zero (i.e., the time of valve closure) and $L_S$ is the interval over which to smoothly transition to zero. For example,  the velocity $\bv_0(t)$ in Equation~\eqref{eq:voft} can be smoothly brought to zero over a discrete interval of $N_S$ timesteps of size $\Delta t$ (corresponding to an inteval $L_S=N_S\Delta t$) by multiplying the velocity by the smoothing function as
\begin{eqnarray}\label{eq:vnew}
\bv_\text{new}(t) = \begin{cases} \bv_0(t), & t\in[0,T_d],\\
\bv_0(T_d)S(t, T_d, N_S\Delta t), & t\in(T_d,T].
\end{cases}
\end{eqnarray}
An illustrative example of the smoothing function $S$ for closure time $t_0 = T_d = 0.5$ s, $\Delta t = 0.005$ s, and $N_S = 15$ is shown in Figure~\ref{fig:Smooth}. One can easily verify that the limit from the left can be found as
\begin{equation}
  \lim_{t\to T_d^-} \bv_\text{new}(t) = \bv_0(T_d),
\end{equation}
and, from the right, as
\begin{eqnarray}
  \lim_{t\to T_d^+} \bv_\text{new}(t) &=&   \lim_{t\to T_d^+} \bv_0(T_d)S(t)\nonumber\\
   &=& \bv_0(T_d)  \lim_{t\to T_d^+}S(t)\\
  &=& \bv_0(T_d),\nonumber
\end{eqnarray}
where we have taken the notational license $S(t) = S(t,T_d,N_S\Delta t).$ Hence $\bv_\text{new}(t)$ is continuous everywhere including at $t=T_d$, i.e., $\bv_\text{new}(t)\subset \mathbb{R} \to \mathbb{R}^3$ is of class ${C}^0([0,T])$. A flowchart summarizing the complete implementation and smooth switching of the 0D LV-elastance model with the $C^\infty$ smoothing employed in this work is presented in Figure~\ref{fig:2}.

\subsection{Algorithmic details}\label{sec:section4}

The following details the numerical algorithms/methods employed for the LB fluid solver and the solid (elastic) solver (which utilizes finite differences for 3D-axisymmetric and finite elements for 3D). Their interactions, described in Section~\ref{sec:IBalg}, can be facilitated by any appropriate fluid-structure algorithm: in this work, an immersed boundary method is used.

\subsubsection{Lattice Boltzmann method\label{sec:LBalg}}

The core algorithm for solving the 3D LB equations consists of a cyclic sequence of sub-steps (where each cycle corresponds to one overall timestep) that is prescribed as:

1. Compute the macroscopic moments $P(\bx,t)$ and $\bm{v}(\bx,t)$ from $f_i(\bx,t)$ via Equations~\eqref{Eqn:rho} and \eqref{Eqn:v}.

2. Obtain the equilibrium distribution $f_i^{eq}(\bx,t)$ from Equation~\eqref{Eqn:feq}.

3. Perform collision (relaxation) and streaming (propagation) to update $f_i(\bx,t)$ via Equation~\eqref{Eqn:fi}.\\
Further details about the implementation of the 3D LB method can be found elsewhere~\cite{wei2020significance,boyd2007analysis, he1997lattice, timm2016lattice}.

For some of the performance studies discussed in Section~\ref{sec:performance}, a 3D-axisymmetric LB model is implemented, where an incompressible D2Q9 BGK model is used to derive an axisymmetric configuration. In such a formulation, for the pseudo-Cartesian coordinates $\bx = (x,r)$ that describe 3D axisymmetric flow, Equation~\eqref{Eqn:fi} can be transformed into
\begin{eqnarray}
\label{Eqn:Afi}
f_i(\boldsymbol{x}+\boldsymbol{e}_i\Delta t,t+\Delta t)-f_i(\boldsymbol{x},t)=-\frac{1}{\tau}[f_i(\boldsymbol{x},t)
-f_i^{eq}(\boldsymbol{x},t)]+\Delta t F_i(\boldsymbol{x},t) + H_i(\boldsymbol{x},t),
\end{eqnarray}
where the a source term $H_i(\boldsymbol{x},t)=\Delta t h^{(1)}_i(\boldsymbol{x},t) + \Delta t^2 h^{(2)}_i(\boldsymbol{x},t)$  is incorporated for microscropic evaluation as~\cite{lee2005axisymmetric}
\begin{eqnarray}
\label{Eqn:AH1}
h^{(1)}_i=-\frac{\omega_i \rho u_r}{ r} \quad \text{and} \quad
h^{(2)}_i=-\omega_i\frac{ 3\nu }{r}\left[\partial_y P + \rho \partial_x u_x u_r +\rho\partial_r u_r u_r + \rho(\partial_r u_x - \partial_x u_r) e_{ix}\right].
\end{eqnarray}

With the inflow conditions given by the complex hybrid ODE-Dirichlet system of the LV-elastance model (Section~\ref{sec:governingLV}), the corresponding outflow conditions can be physiologically modeled by any suitable boundary condition that accounts for downstream physiological effects including the effective compliance, resistance and wave reflection of the truncated vasculature (to approximate the effect of the eliminated peripheral vessels). This is reasonably captured in this work using the extension outflow boundary tube model consisting of an elastic tube terminated in a rigid contraction~\cite{pahlevan2011physiologically}. Such a model has been successfully utilized for hemodynamics studies~\cite{weiPRF,aghilinejad2021effects,pahlevan2014intrinsic,kang2019accuracy}. Other models can be used, including lumped parameter Windkessel ODEs~\cite{amlanipahlevan,westerhof1969analog,vignon2004outflow} (which can be easily implemented using the same direct 0D-3D coupling introduced earlier).

\subsubsection{Fluid-structure interactions (FSI)\label{sec:IBalg}}

For all 3D simulations of this work (including the physiological example study of Section~\ref{sec:casestudy}), the solid deformation given by Equation (\ref{Eqn:structure}) is numerically simulated by the nonlinear finite element method (FEM) solver of previous study \cite{doyle2013nonlinear}, where the the large-displacement and small-strain deformation problems are handled by co-rotational schemes.
The numerical strategy has been successfully implemented in previous works for resolving a wide range of fluid-structure interaction (FSI) problems incorporating elastic structures~\cite{hua2014dynamics,dai2012dynamic, tang2015dynamics, huang2018coupling}. Briefly, such a method uses three-node triangular elements to describe the deformation using six degrees of freedom (three displacement components and three angles of rotation)~\cite{batoz1980study}. An iterative strategy is then used for the time integration of the subsequent nonlinear systems of algebraic equations in order to ensure second-order accuracy. A further detailed description of the particular finite element method employed in this work can be found elsewhere~\cite{doyle2013nonlinear}.

{
For the 3D-axisymmetric performance studies included in Section~\ref{sec:performance}, a self-implemented staggered grid finite difference (SGFD) methodology~\cite{huang2007simulation} is employed in the Lagrangian coordinate system (where $\bm{s} = s \in \mathbb{R}$ is the arc length), where only the tension force given by
\begin{equation}\label{eq:tension}
\bm{T}=Eh\left(\sqrt{\frac{\partial \bm{X}}{\partial s}\cdot\frac{\partial \bm{X}}{\partial s}}-1\right)\end{equation}
is defined on the interface (the displacement variable $\bm{X}(s,t)$, for example, is defined on all the nodes). {The solid deformation governed by Equation~\eqref{Eqn:structure} is subsequently solved  by such a finite difference methodology in a strong form~\cite{huang2007simulation,connell2007flapping,huang2010three}}. That is, for an arbitrary variable, the central, downwind and upwind difference approximations to the first-order derivatives, are given by
\begin{equation}
  \begin{cases}
D^0_s \bm{X} & =(\bm{X}(s+\Delta s/2)-\bm{X}(s-\Delta s/2))/\Delta s,\\
D^+_s \bm{X} & =(\bm{X}(s+\Delta s)-\bm{X}(s))/\Delta s,\\
D^-_s \bm{X} & =(\bm{X}(s)-\bm{X}(s-\Delta s))/\Delta s,
  \end{cases}
\end{equation}
such that the corresponding second-order central difference approximation can be defined as
\begin{equation}
\label{2nd order}
D^+_s D^-_s \bm{X}=(\bm{X}(s+\Delta s)-2\bm{X}(s)+\bm{X}(s-\Delta s))/\Delta s^2,
\end{equation}
where the same difference approximation is applied for the time derivative. The tension force term of Equation~\eqref{Eqn:structure} (given by Equation~\eqref{eq:tension}) is hence approximated as
\begin{equation}
\label{tension}
D_s(\bm{T}D_s \bm{X})=D^0_s(\bm{T}D^0_s \bm{X})=\frac{\bm{T}D^0_s \bm{X}_{s+\Delta s/2}-\bm{T}D^0_s \bm{X}_{s-\Delta s/2}}{\Delta s}.
\end{equation}
Similarly. the bending force term in Equation (\ref{Eqn:structure}) can be approximated as
\begin{equation}
\label{bending}
-D_{ss}(EI D_{ss} \bm{X})=-D^+_sD^-_s(EI D^+_sD^-_s \bm{X})=-EI\frac{D^+_sD^-_s \bm{X}_{s+\Delta s}-2D^+_sD^-_s \bm{X}_{s}+D^+_sD^-_s \bm{X}_{s-\Delta s}}{\Delta s^2}.
\end{equation}

For coupling the fluid and solid systems, any suitable FSI coupling strategy can be employed. For this particular work, the immersed boundary (IB) method~\cite{goldstein1993modeling} is used to couple the LB method of the fluid with the 3D FEM (or 3D-axisymmetric FDM) of the solid~\cite{huang2007simulation, hua2014dynamics}.
This method has been extensively used to simulate FSI problems in cardiovascular biomechanics \cite{weiPRF,peskin2002immersed,mittal2005immersed,seo2013multiphysics}. The body force term $\bm{b}(\bx,t)$ in Equation~\eqref{Eqn:f} is used as an interaction force between the fluid and the boundary in order to enforce the no-slip velocity boundary condition at the FSI interface. The Lagrangian force between the fluid and structure, $\bm{B}(\bm{s},t)$, is then calculated by a penalty scheme~\cite{goldstein1993modeling} using the expression given by
\begin{equation}
\label{Eqn:LagForce}
\boldsymbol{B}(\bm{s},t)=\alpha\left(\int^t_0\left(\boldsymbol{V}_f(\bm{s},t^\prime)-\boldsymbol{V}_s(\bm{s},t^\prime)\right)\mathrm{d}t^\prime\right) +\beta\left(\boldsymbol{V}_f(\bm{s},t)-\boldsymbol{V}_s(\bm{s},t)\right),
\end{equation}
where $\alpha\in\mathbb{R}$ and $\beta\in\mathbb{R}$ are negative penalty parameters (adopted in this work from previous studies~\cite{hua2014dynamics, tang2015dynamics, ye2017two, zhang2017free, huang2018coupling}); $\boldsymbol{V}_s(\bm{s},t)=\partial \boldsymbol{X}/\partial t$ is the velocity of Lagrangian material point of the solid wall; and $\boldsymbol{V}_f(\bm{s},t)$ is the fluid velocity at the position $\boldsymbol{X}$. The latter can be obtained through transforming the Eulerian fluid velocity $\boldsymbol{v}(\boldsymbol{x},t)$ into Lagrangian coordinates via the integral
\begin{eqnarray}
\label{Eqn:interpolation}
\boldsymbol{V}_f(\bm{s},t)=\int\boldsymbol{v}(\boldsymbol{x},t)\delta(\boldsymbol{x}-\boldsymbol{X}(\bm{s},t)){\rm d}\boldsymbol{x},
\end{eqnarray}
where $\delta(\boldsymbol{x}-\boldsymbol{X}(\bm{s},t)$ is a Dirac delta function. The body force $\bm{b}(\bm{x},t)$ in Eulerian coordinates can be calculated from the corresponding Lagrangian body force via the expression
\begin{eqnarray}
\label{Eqn:EulForce}
\boldsymbol{b}(\boldsymbol{x},t)=-\int\boldsymbol{B}(\bm{s},t)\delta(\boldsymbol{x}-\boldsymbol{X}(\bm{s},t)){\rm d}\bm{s}.
\end{eqnarray}
 The Lagrangian interaction force $\bm{B}$ can be explicitly obtained by the penalty IB strategy described above. Such a formulation of the IB numerical strategy has been successfully applied to a wide range of FSI problems~\cite{hua2014dynamics, tang2015dynamics, ye2017two, zhang2017free}, including those governed by the dynamics of fluid flow over a circular flexible plate~\cite{hua2014dynamics} and an inverted flexible plate~\cite{tang2015dynamics}. The overall numerical algorithm for solving the complete FSI system is summarized in the block-diagram of Figure~\ref{fig:IB-FSI}, and its corresponding pseudocode implementation is provided in Algorithm~\ref{alg:solver}. At the start of each numerical simulation, the 0D and 3D domains can be initialized with $U_0=0$, $Q_0=0$ and an arbitrary $P_0\neq 0$, respectively.

\begin{figure}
\centering
\includegraphics[width=.7275\textwidth]{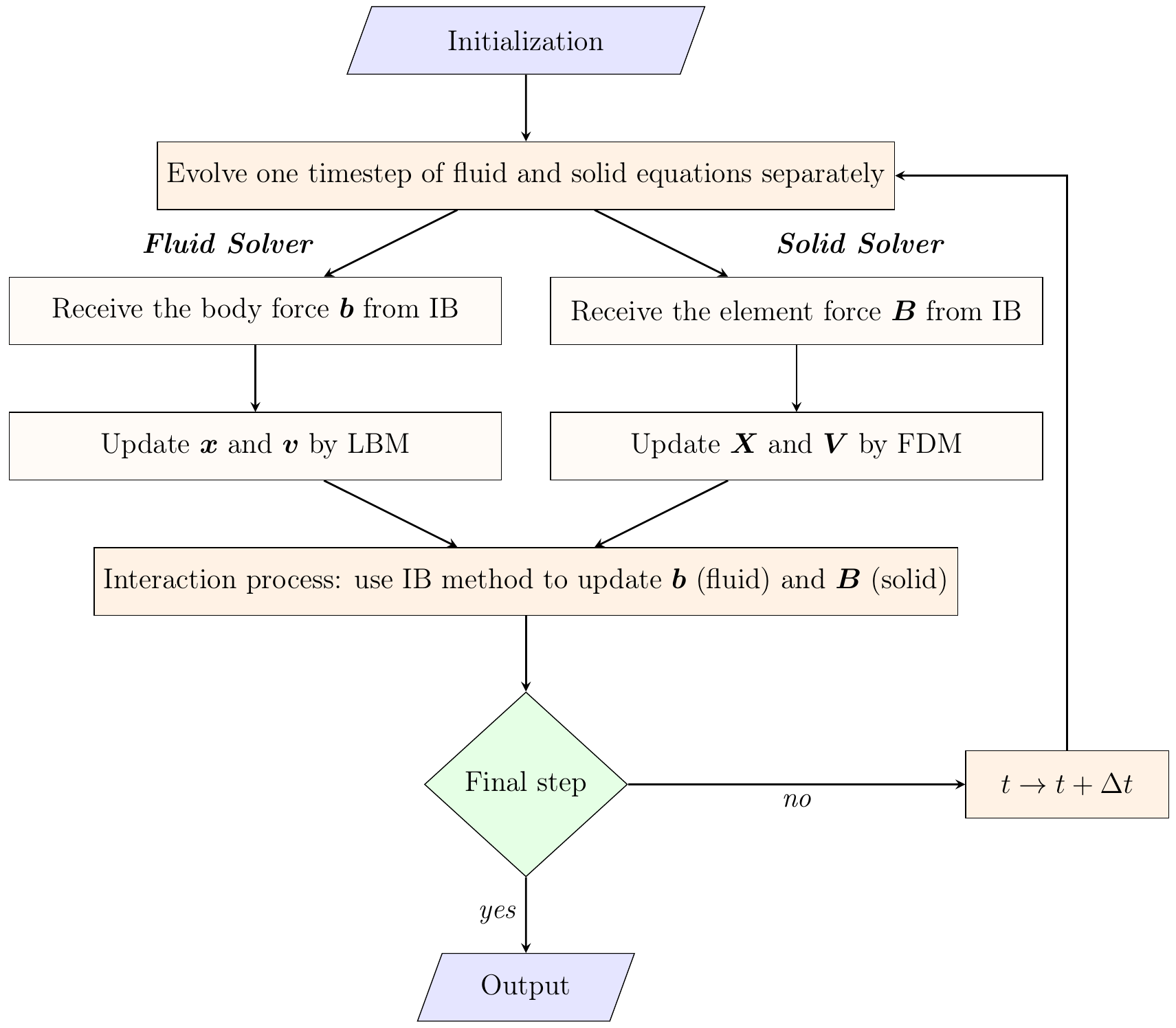}% Here is how to import EPS art
\caption{\label{fig:IB-FSI} A fluid-structure interaction (FSI) procedure facilitated by the immersed boundary (IB) method.}
\end{figure}
\begin{algorithm}
\vspace{2mm}
 \textbf{Input} fluid material parameters $\text{Re}, \text{W}_o, \mu$\\
 \textbf{Input} solid material parameters $Eh, EI$ \\
 \textbf{Input} characteristic domain parameters $D, L_{x_i}$\\
 \textbf{Input} LV-elastance parameters $C_v(t), {\partial C_v}/{\partial t}(t)$\\
 \textbf{Input} numerical discretizations $\Delta x_i, \Delta s_i, \Delta t$\\
 \textbf{Input} total number of cardiac cycles to simulate ($\implies$ total time $T_\text{max}$ )\\
\begin{algorithmic}[1]
  % \STATE Input system parameters ${L}, {C}, {R}$
  % \STATE Input uniform grid $\{x=0,x_1,...,x = \ell\}$
  % \STATE Input number of cardiac cycles to simulate \COMMENT{\small$\implies$ final time $T_f$}\\

  \STATE  Initialize the pressure $P$ and velocity $\boldsymbol{v}$ for the fluid \COMMENT{\small initial time $t=0$}
  \STATE  Initalize $\boldsymbol{X}$ for the solid structure position \COMMENT{\small initial time $t=0$}
\WHILE{$t<T_\text{max}$}
\STATE Obtain macroscopic fields $P$ and $\boldsymbol{v}$ from distribution functions $f_i$ \COMMENT{\small via Eq.~\eqref{Eqn:rho} \& Eq. \eqref{Eqn:v}}
% \STATE  Interpolate the fluid velocity to obtain $\partial \boldsymbol{X}/\partial t$ at the immersed boundary
\STATE Compute Lagrangian interaction force $\bB(\bs,t+\Delta t)$ on $\partial \Omega_1$ via IB \COMMENT{\small via Eq.~\eqref{Eqn:LagForce}}
\STATE Compute the corresponding body force $\bb(\bx,t+\Delta t)$ in $\Omega$ via IB \COMMENT{\small via Eq.~\eqref{Eqn:EulForce}}
\STATE Evolve ODE (i.e., 0D-model) to $t+\Delta t$
\STATE Couple ODE with LBM to evolve $f_i, f_{eq}$ to $t+\Delta t$ on $\partial \Omega_2$
\COMMENT{\small via Fig.~\ref{fig:eout}}
\STATE Perform collision \& streaming (LBM) to evolve $f_i, f_{eq}$ to $t+\Delta t$ in $\Omega$
\COMMENT{\small via Eq.~\eqref{Eqn:fi}}
\STATE Evolve solid wall position $\bX$ to $t+\Delta t$ on $\partial \Omega_1$ via FDM/FEM
\COMMENT{\small via Eq.~\eqref{Eqn:structure}}
\STATE Update the macroscopic variables $P,\bv$ to $t+\Delta t$ in $\overline{\Omega}$
\STATE $t\rightarrow t+\Delta t$
\ENDWHILE
\end{algorithmic}
\hspace*{\algorithmicindent}\\
 \textbf{Output}  the numerical solutions for $\bv(\bx,t), P(\bx,t), \bX(\bs,t)$ on $\overline{\Omega}\times[0,T_\text{max}]$
 \vspace{2mm}
%\hrule
\caption{Summary of a complete 3D IB-LBM solver with 0D coupling.}
\label{alg:solver}
\end{algorithm}

\section{Results and Discussion}\label{sec:results}

\subsection{Performance evaluation}\label{sec:performance}

This section presents a series of performance studies, based on benchmark cases and manufactured solutions, that validate the implementation and convergence of the solid, fluid and interaction components of the complete solver (including the LV-elastance model). Firstly, the fluid solver is validated by non-oscillatory and oscillatory cases of a flow around a cylinder. The solid solver is then validated through a classical case of a hanging elastic filament as well as the method of manufactured solutions (MMS).

\subsubsection{{Fluid solver: steady and oscillating cylinders}}

In order to validate the present LBM solver and its coupling with the IB method for solid wall boundaries, two cases are considered: a uniform flow passing a non-oscillating cylinder and uniform flow passing a transversely oscillating cylinder. In those cases the motion of the solid wall is prescribed (rigidly), and hence the problem becomes a one-way fluid-structure coupling such that the solid position of the center of the cylinder is given by
\begin{equation}
  \bm{X}_c=(-10D, A_m \cos(\omega_e t))^T,
\end{equation}
 where $D$ is the cylinder diameter; and $A_m \in \mathbb{R}$, $\omega_e\in\mathbb{R}$ are the constant oscillation amplitude and frequency, respectively. In the cases considered in this section, the Reynolds number is defined as $\text{Re} = vD/\nu=185$ for velocity $v$ (m/s), fluid viscosity $\nu$ and $D=1$ m. Defining $\omega_0 = 0.4\pi$ radians as the natural shedding frequency for a stationary cylinder, the computation is performed with the parameters $A_m = 0.2$ m, and two oscillating cases $\omega_e/\omega_0=0.9$ and $\omega_e/\omega_0=1.1$ (note that $\omega_e/\omega_0=0.0$ for a non-oscillating cylinder). These test cases and their corresponding parameters are adopted from the benchmark cases proposed by previous study~\cite{guilmineau2002numerical}. Figures~\ref{fig:2D} (left) illustrates the computational domain for both steady and moving cases.

\begin{figure}
\centering
\hfill\includegraphics[width=0.35\textwidth,valign=m]{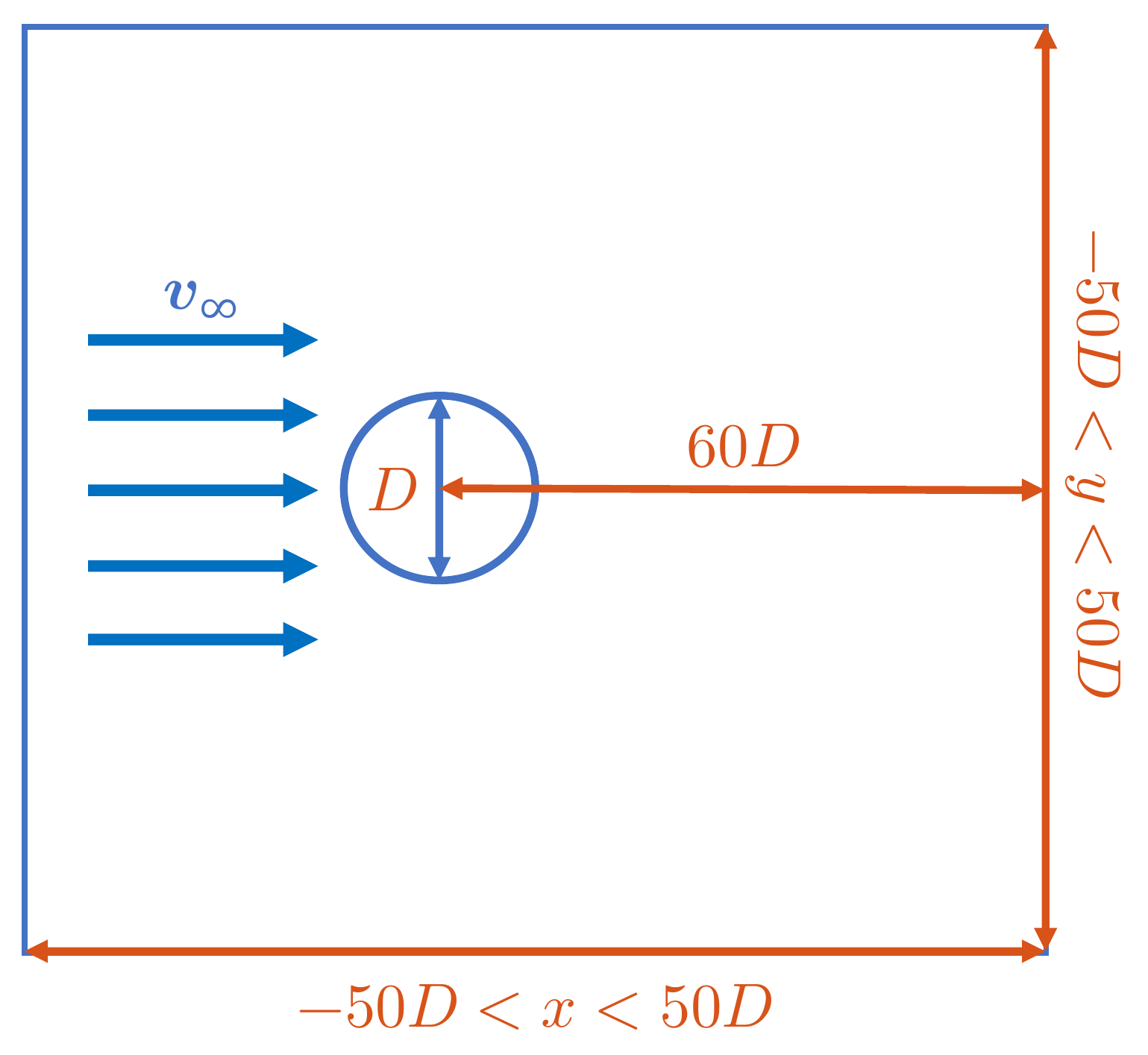}\hfill\includegraphics[width=0.485\textwidth, valign=m]{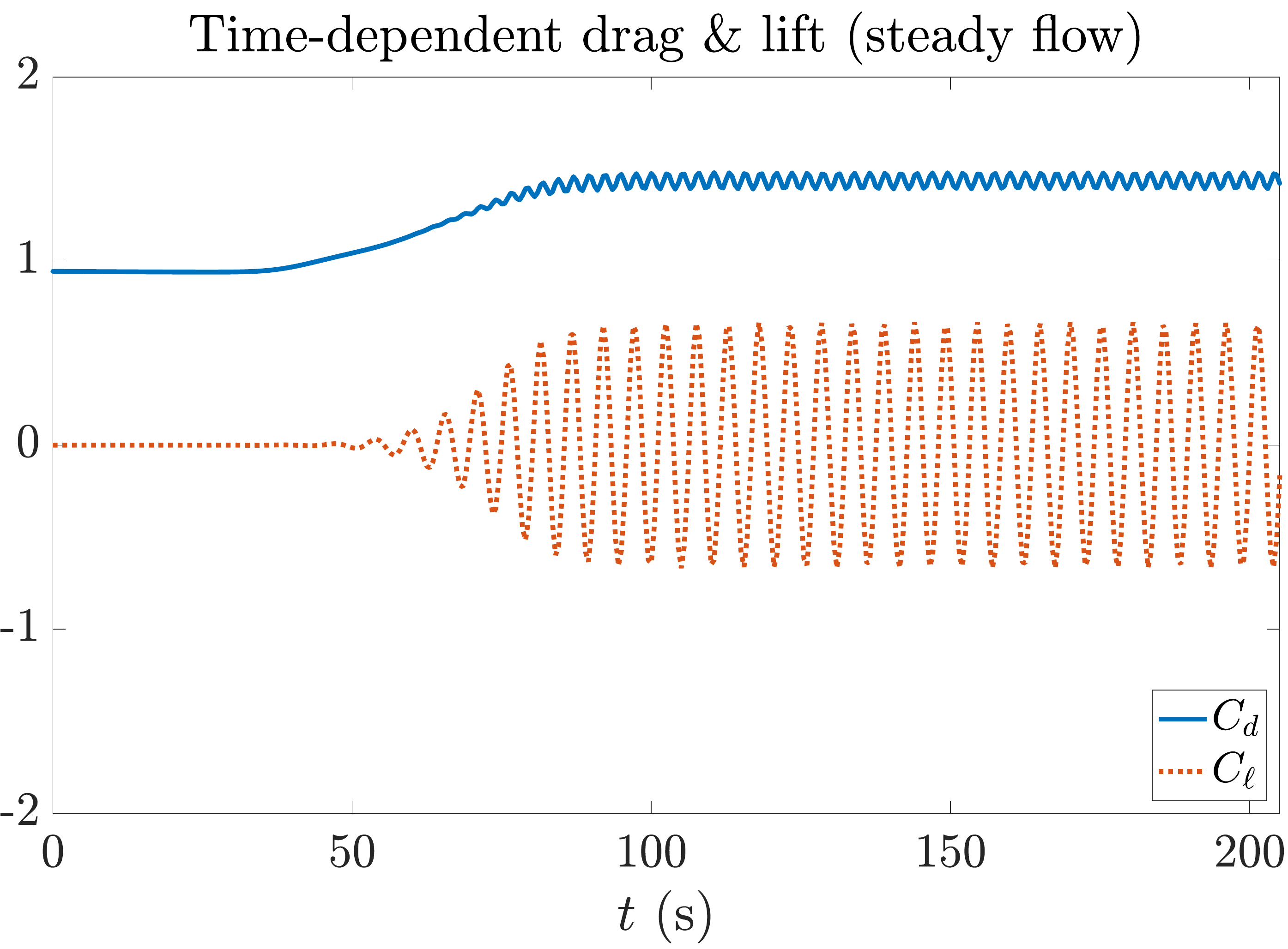}% Here is how to import EPS art
\caption{\label{fig:2D} Numerical validation of the steady flow (non-moving cylinder) test case. (Left) Diagram of the 2D computational domain. (Right) The corresponding temporal evolution of drag (solid blue) and lift (dotted red) coefficients at $\text{Re}=185$, demonstrating excellent agreement with previous studies~\cite{guilmineau2002numerical,kim2006immersed}.}
\end{figure}

 Figure~\ref{fig:2D} (right) presents the corresponding time evolution of the drag and lift coefficients $C_d$, $C_l$ of the steady case as simulated by our solver up to a final time of $t = 400$s. The numerical discretization corresponds to  $\Delta x = 1/32 D, \Delta t = 1/100 s$ (where $D=1$ m). Using the same discretization and computational domain, Figure~\ref{fig:V_F1} additionally presents the time evolution of the drag and lift coefficients of the moving (oscillating) cylinder case for  $\omega_e/\omega_0=0.9$ (left) and $\omega_e/\omega_0=1.1$ (right). All of these cases (steady and moving) are in excellent agreement with those results presented in previous studies~\cite{guilmineau2002numerical,kim2006immersed}.

\begin{figure}
\centering
\includegraphics[width=0.485\textwidth, valign=m]{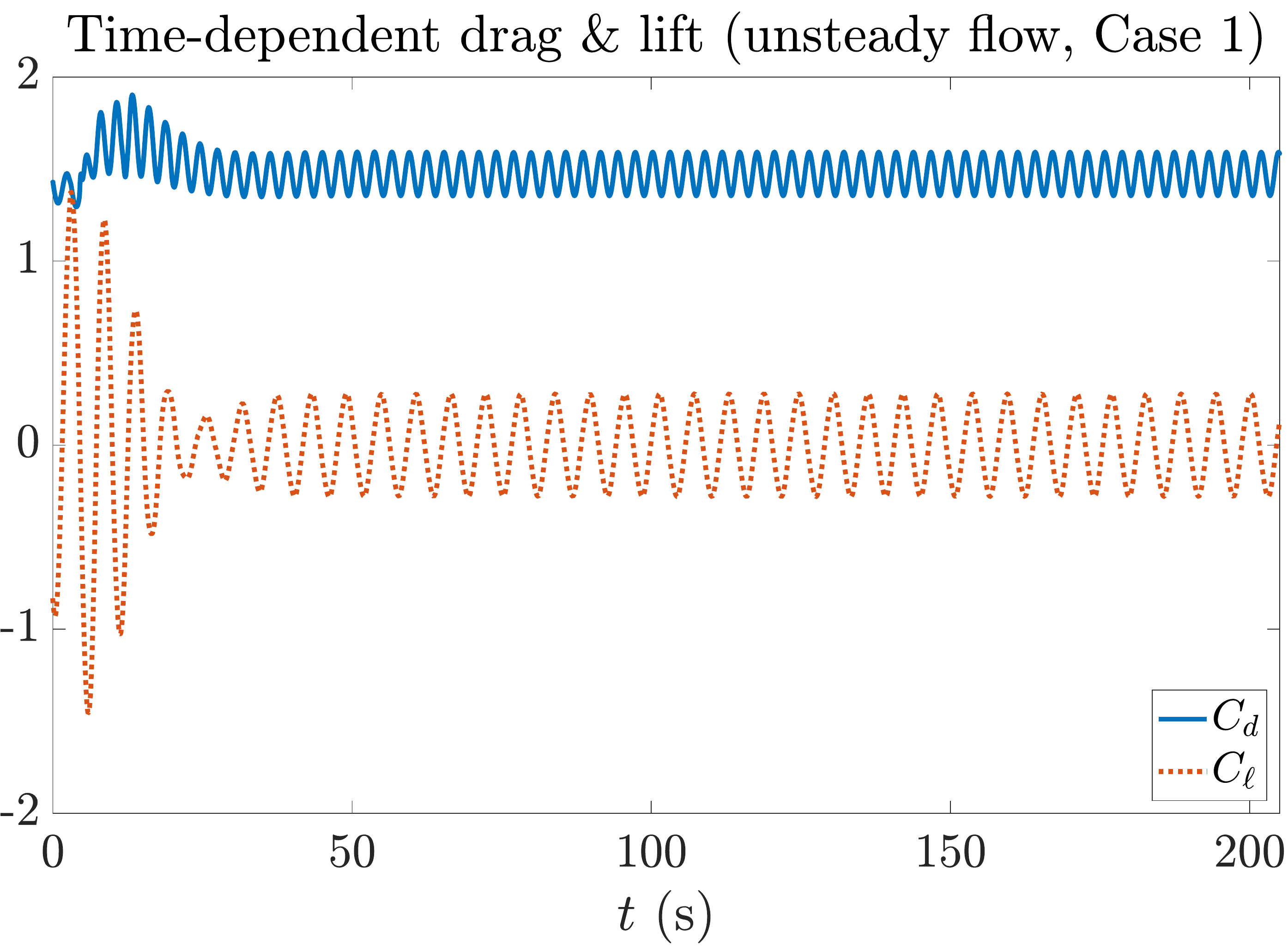}\hfill% Here is how to import EPS art
\includegraphics[width=0.485\textwidth, valign=m]{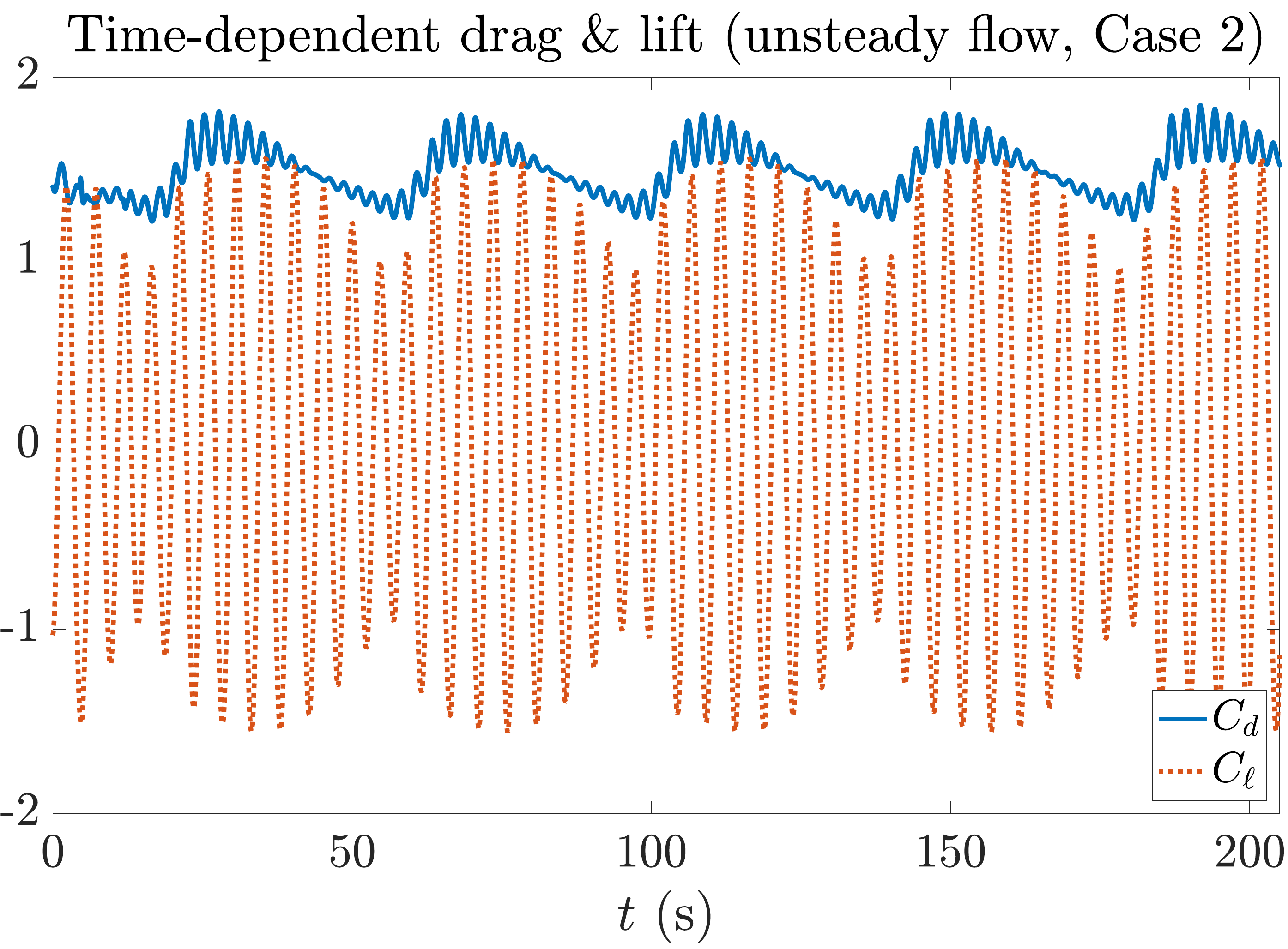}% Here is how to import EPS art
\caption{\label{fig:V_F1}
Numerical validation of the unsteady flow (oscillating cylinder) test case.
(Left) The temporal evolution of drag (solid blue) and lift (dotted red) coefficients at $\text{Re} = 185, A_m/D = 0.2$ and $f_e/f_0=0.9$. (Right) The temporal evolution of drag and lift coefficients at $\text{Re} = 185, A_m/D = 0.2, f_e/f_0=1.1$. Both simulated cases are in excellent agreement with results provided by previous studies~\cite{guilmineau2002numerical,kim2006immersed}.}
\end{figure}

\subsubsection{{Solid solver: filament under gravity and manufactured solutions}}

In order to validate the elastic (solid) component,  two case are considered: an elastic filament under gravity and a manufactured solution (i.e., a given right-hand-side).

The first case considers a hanging filament of length $L=1$ m without an ambient fluid (in order to independently test the solid solver alone), as proposed by Huang et at.~\cite{huang2007simulation}. Adopting the same arbitrary units, the filament is initially held stationary at an angle $k=0.1\pi$ radians from the vertical, where the physical parameters correspond to a Froude number $Fr = v/\sqrt{gL} = 10.0$ ($g$ is the gravitation constant) and a bending rigidity $EI=0.01$ Pa m$^3$. Snapshots of the simulated positions of the filament over a time period of 0.8 s are illustrated in Figure~\ref{fig:V_S1}, where a spatial discretization of $N = 100$ elements is utilized together with a timestep of $\Delta t = 0.0001$ s. Again, the solutions produced by the solver of this work are in excellent agreement with those results presented by Huang et al.~\cite{huang2007simulation}.

\begin{figure}
\centering
\includegraphics[width=.485\textwidth,valign=m]{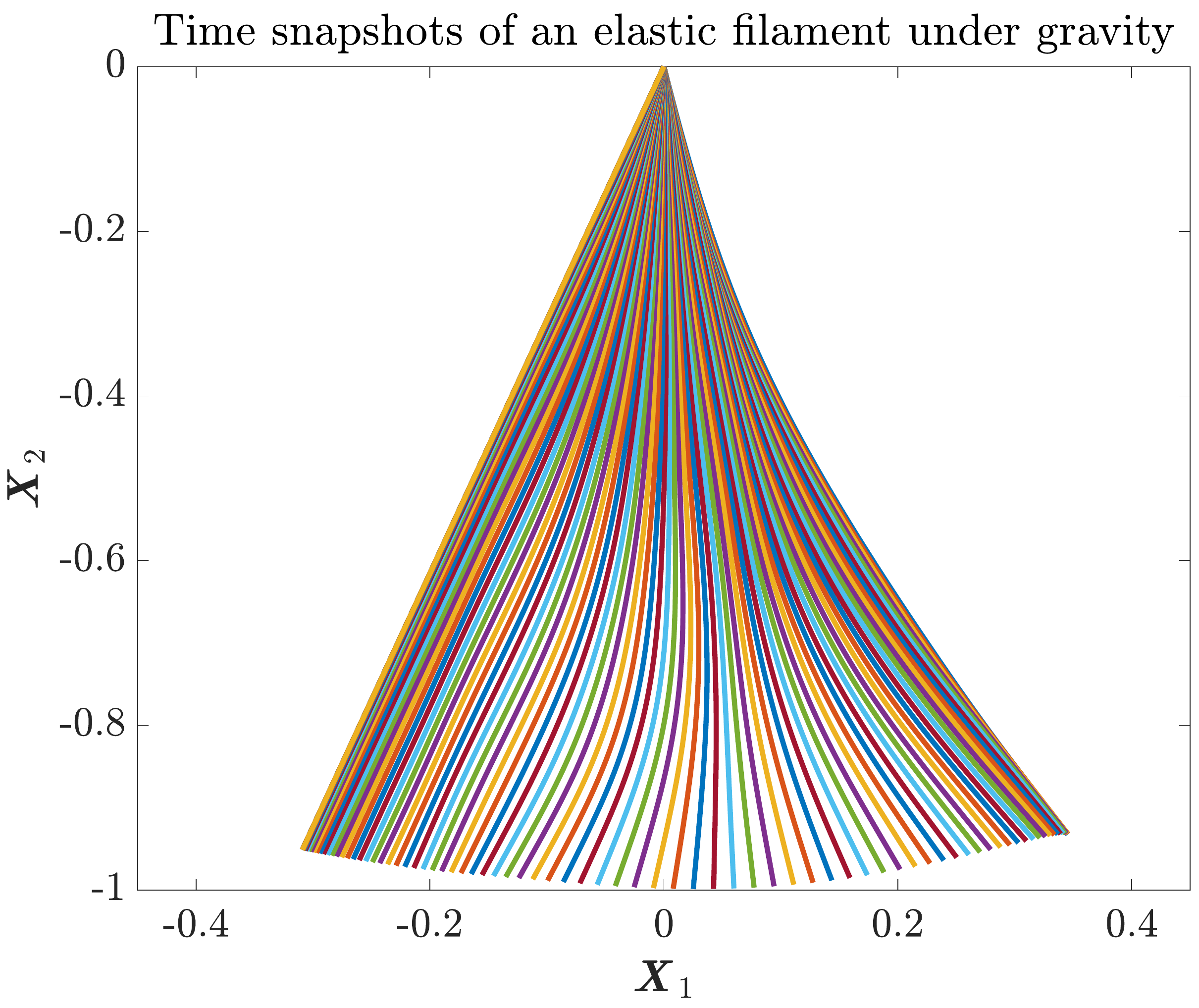}% Here is how to import EPS art
\caption{\label{fig:V_S1} Numerical validation of the solid solver using an elastic filament. Superposition of simulated filament positions (from left to right) at successive times, demonstrating excellent agreement with positions provided by Huang et al.~\cite{huang2007simulation}.}
\end{figure}

In order to further validate the implementation of the solid solver (in particular, it's numerical accuracy), the method of manufactured solutions (MMS) is additionally employed. Such a verification procedure has been extensively used for validating other hemodynamics solvers~\cite{amlanipahlevan,raghu2011verification,raghu2011comparative}.  {In MMS, one proposes a closed-form smooth solution (i.e., arbitrary movement) and subsequently derives (algebraically) the corresponding right-hand forcing terms and boundary conditions in order to render the postulated function to be an exact solution of the solid equations. Here, we postulate a given displacement function $\bm{X}(s,t)$ as
\begin{equation}
  \label{Eqn:MMSsolid}
  \bm{X}(s,t)=(s, A \cos(\pi s)\sin(\pi t))^T,
\end{equation}
{where $A=0.01$ {m} is the maximum amplitude of displacement in the vertical component. The solid structure considered is again an elastic filament of length {$L=1$ m} with elastic material parameters corresponding to a Young's modulus of {$E=0.5$ {MPa} and a thickness of $h=1$ mm}.} Employing the same numerical discretization as in the gravity case, Figure~\ref{fig:V_S3} (left) presents snapshots in time of the simulated filament positions for both numerical (solid lines) and analytical values (dashed lines).

\begin{figure}
\centering
\includegraphics[width=.485\textwidth,valign=m]{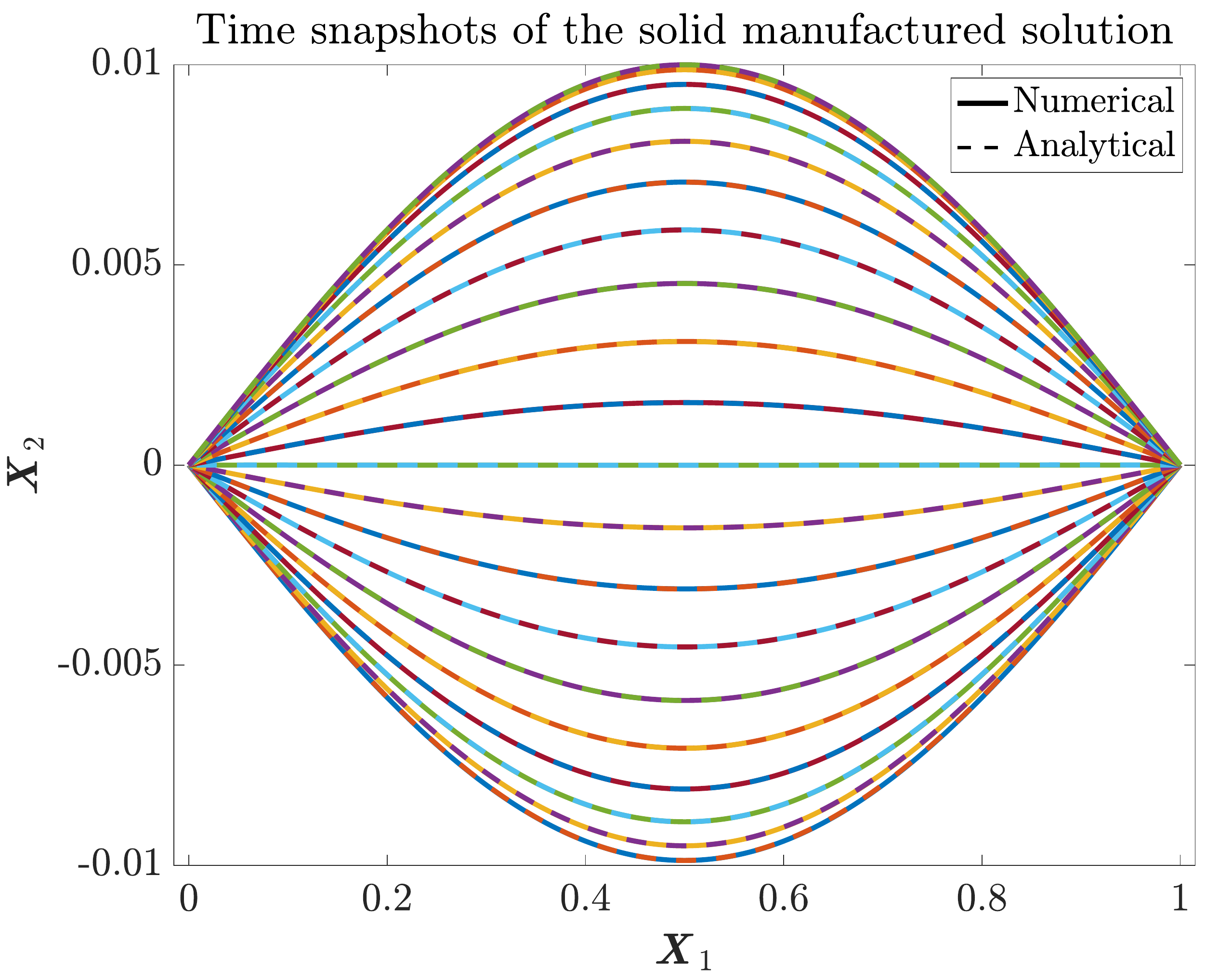}\hfill
\includegraphics[width=.485\textwidth,valign=m]{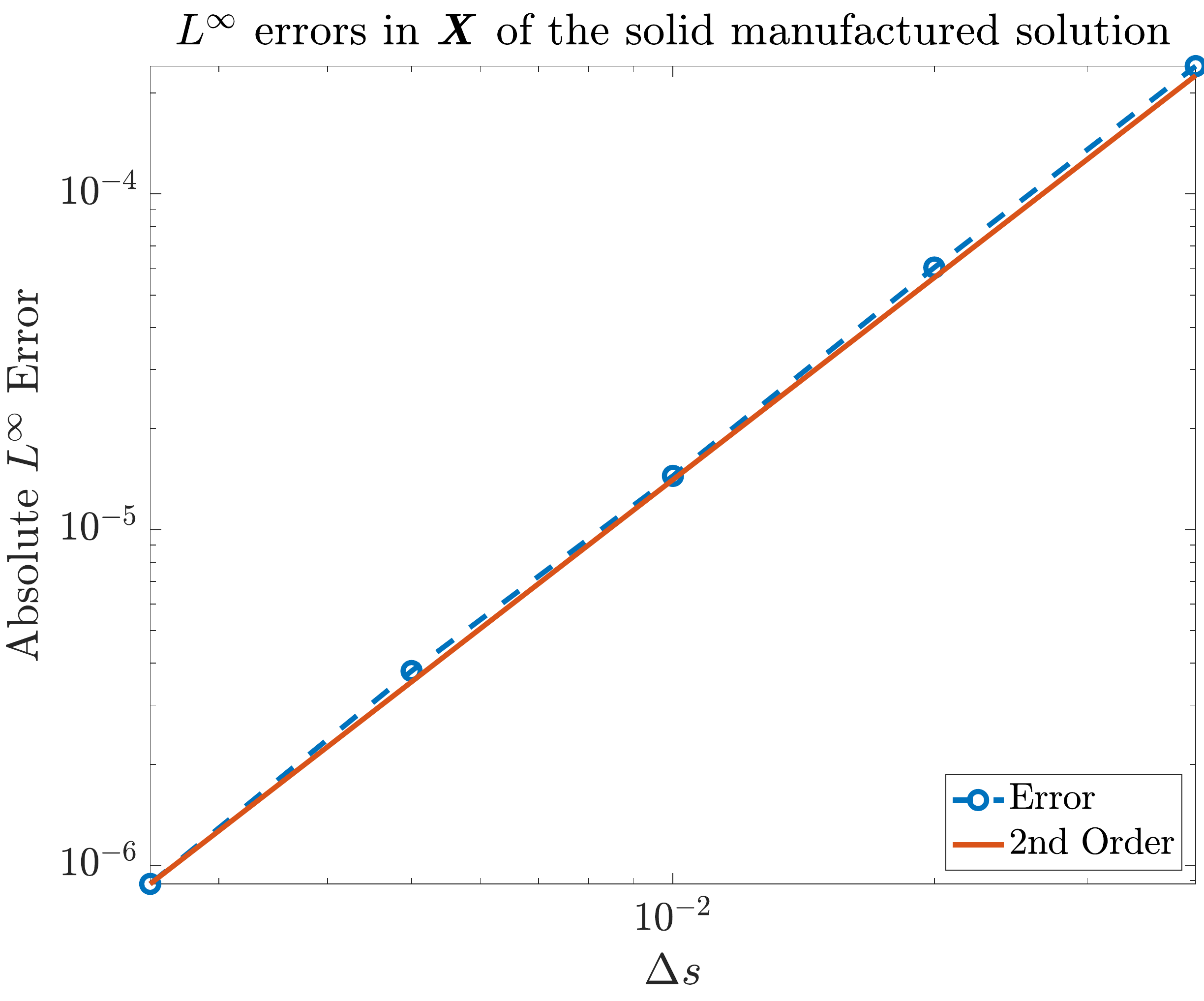}% Here is how to import EPS art
\caption{\label{fig:V_S3} (Left) Snapshots in time of a manufactured solution (Equation~\eqref{Eqn:MMSsolid}) of both the exact values (dashed lines) and those produced by our numerical simulations (solid lines). (Right) The corresponding $L^\infty$ errors between the numerical and analytical solutions, which demonstrate the expected second-order accuracy of the solid solver.}
\end{figure}

Using successive discretization sizes that are integer multiples of the coarsest one used ($N=25$ elements), the simulation is advanced for $100,000$ timesteps in all cases at a fine time-step size of $\Delta t=1\cdot10^{-5}$ s (in order to ensure that errors are dominated by the spatial discretization). The maximum absolute errors between simulated displacements and the exact manufactured solution of Equation~\eqref{Eqn:MMSsolid}, over all space and for all timesteps, are presented in Figure~\ref{fig:V_S3} (right). The overlaid slopes in the plots illustrate the expected second-order of accuracy for the elastic (solid) solver employed in this paper (hence verifying its implementation).

\subsubsection{{The complete FSI solver coupled to the 0D LV-elastance heart model}}

In order to verify the complete FSI solver coupled to a 0D hybrid ODE-Dirichlet heart model, one can consider the axisymmetric straight aorta configuration (of length $25D$) presented in Figure~\ref{fig:Sche}. Adopting parameters of the LV-elastance hybrid ODE-dirichlet model from previous work~\cite{amlanipahlevan} corresponding to an end-systolic LV elastance $E_{es} = 2.2$mmHg/ml and a $\text{CO} = 4.3$ L/min, Figure~\ref{fig:pin} (left) presents the expected physiologically-accurate pressure profiles at the 0D-3D interface as simulated by the complete solver for discretizations corresponding to $\Delta x = 1/20, 1/32, 1/64, 1/100$ and $1/128$, where physical parameters of $\text{W}_o = 16$, a non-dimensionalized $D = 1$ (corresponding to 24 mm), $\mu = 3.5 \text{ cP}$ and $\rho = 1000 \text{ kg/m}^3$} are employed.  The timestep is fixed again and is taken small enough so that errors are dominated by the spatial discretization. Figure~\ref{fig:pin} (right) presents the corresponding $L^\infty$ errors (relative to the finest solution), where a convergence between first and second order can be observed (and is expected from the second-order nature of the lattice Boltzmann solver and the first-order discretizations of the LV-elastance ODEs). Figure~\ref{fig:qin} (left) and Figure~\ref{fig:pmid} (left) additionally present the simulated physiological flow profiles at the inlet and the pressure profiles at the midpoint of the vessel, respectively. The corresponding $L^\infty$ errors (relative to the finest discretization of  $\Delta x=1/128$) are presented in Figure~\ref{fig:qin} (right) and Figure~\ref{fig:pmid} (right), respectively. As before, one can appreciate the convergence and accuracy as expected from the second-order fluid discretization and the first order LV-elastance ODE time integration.

\begin{figure}
\centering
\includegraphics[width=.65\textwidth,valign=m]{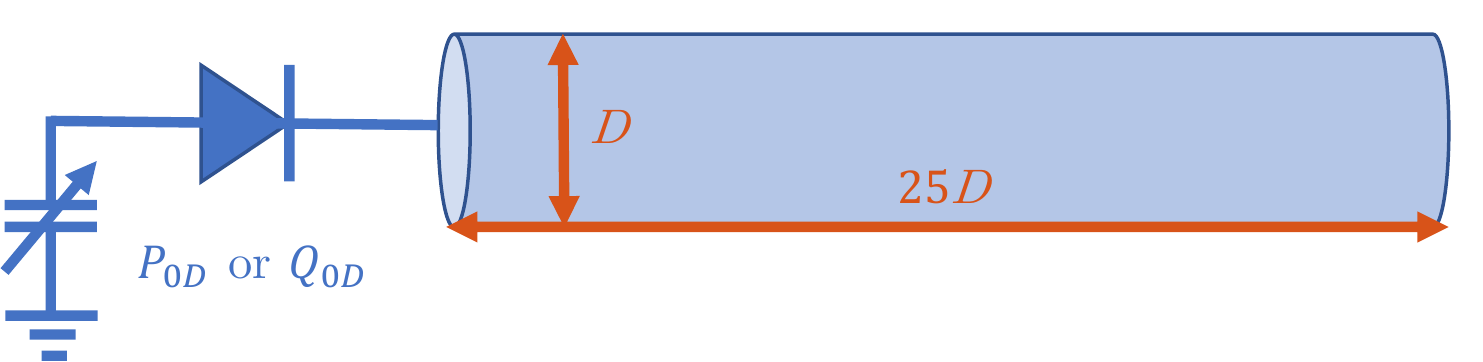}% Here is how to import EPS art
\caption{\label{fig:Sche} Diagram of the simplified straight aorta test case coupled to the LV-elastance model.}
\end{figure}

\begin{figure}
\centering
\includegraphics[width=.485\textwidth,valign=m]{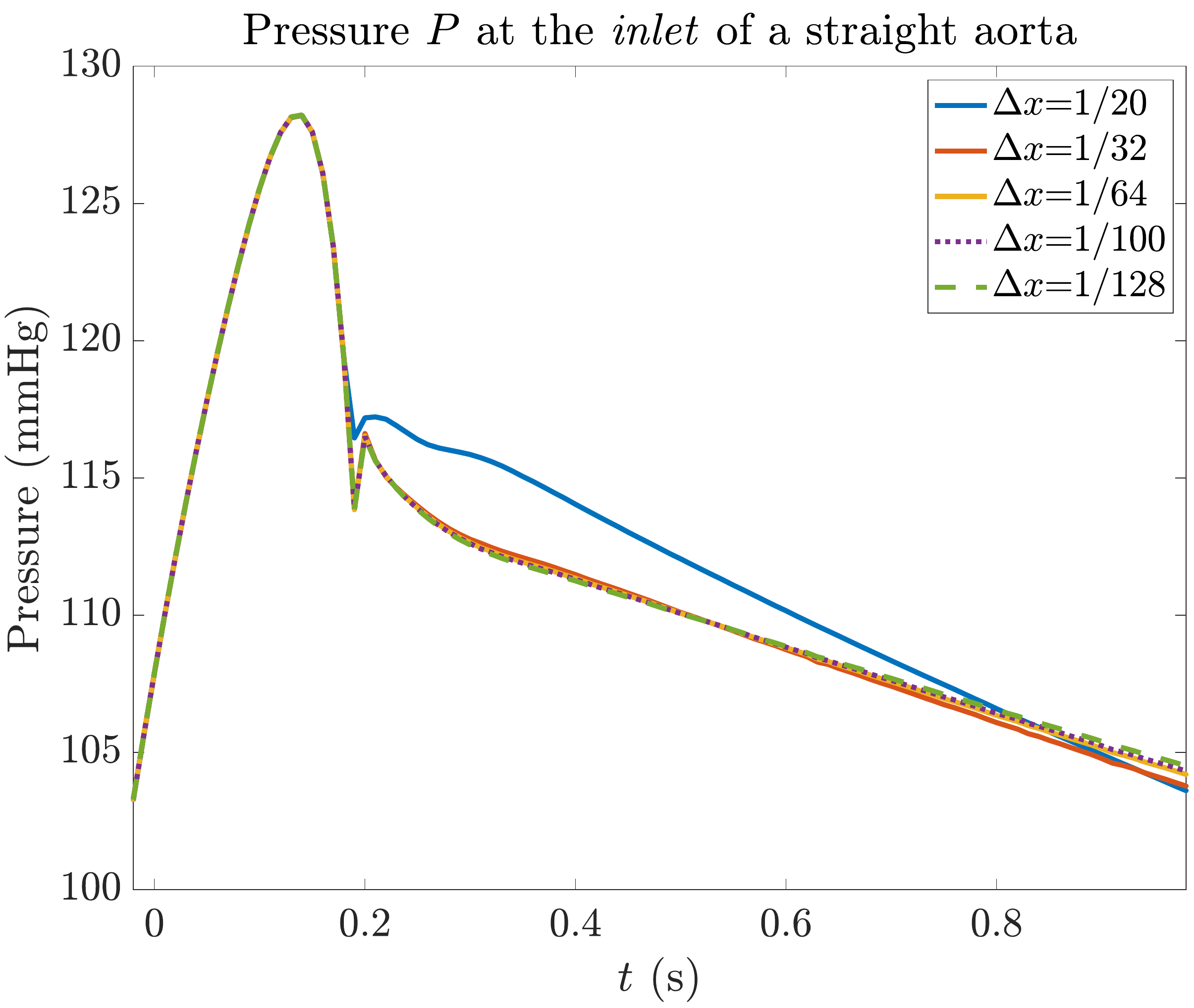}\hfill\includegraphics[width=.485\textwidth,valign=m]{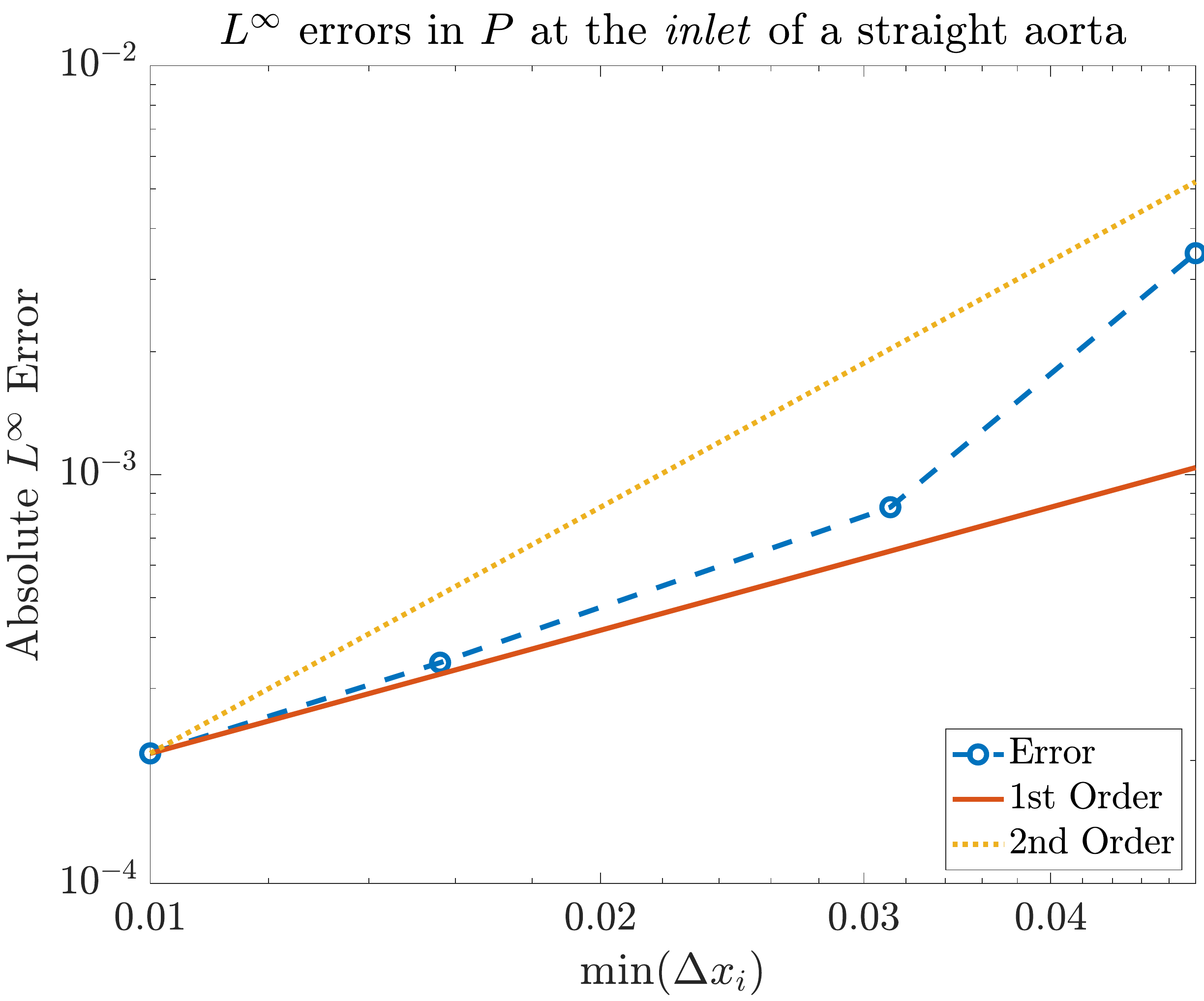}
\caption{\label{fig:pin} (Left) Physiological pressure profiles at the inlet for successively-refined discretizations of a straight aorta as produced by the LV-elastance model. (Right) The corresponding $L^\infty$ errors (relative to the finest solution).}
\end{figure}

\begin{figure}
\centering
\includegraphics[width=.485\textwidth,valign=m]{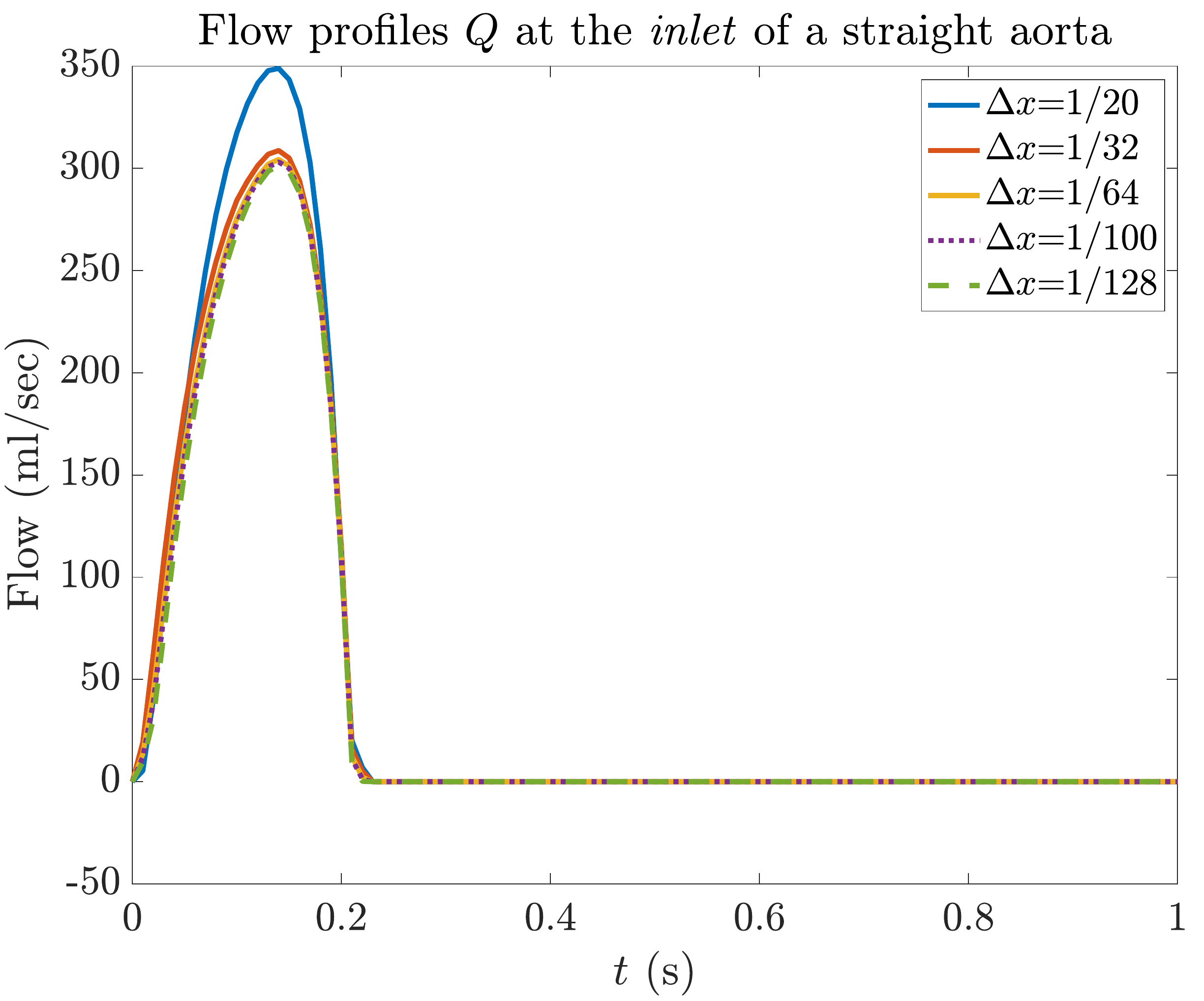}\hfill\includegraphics[width=.485\textwidth,valign=m]{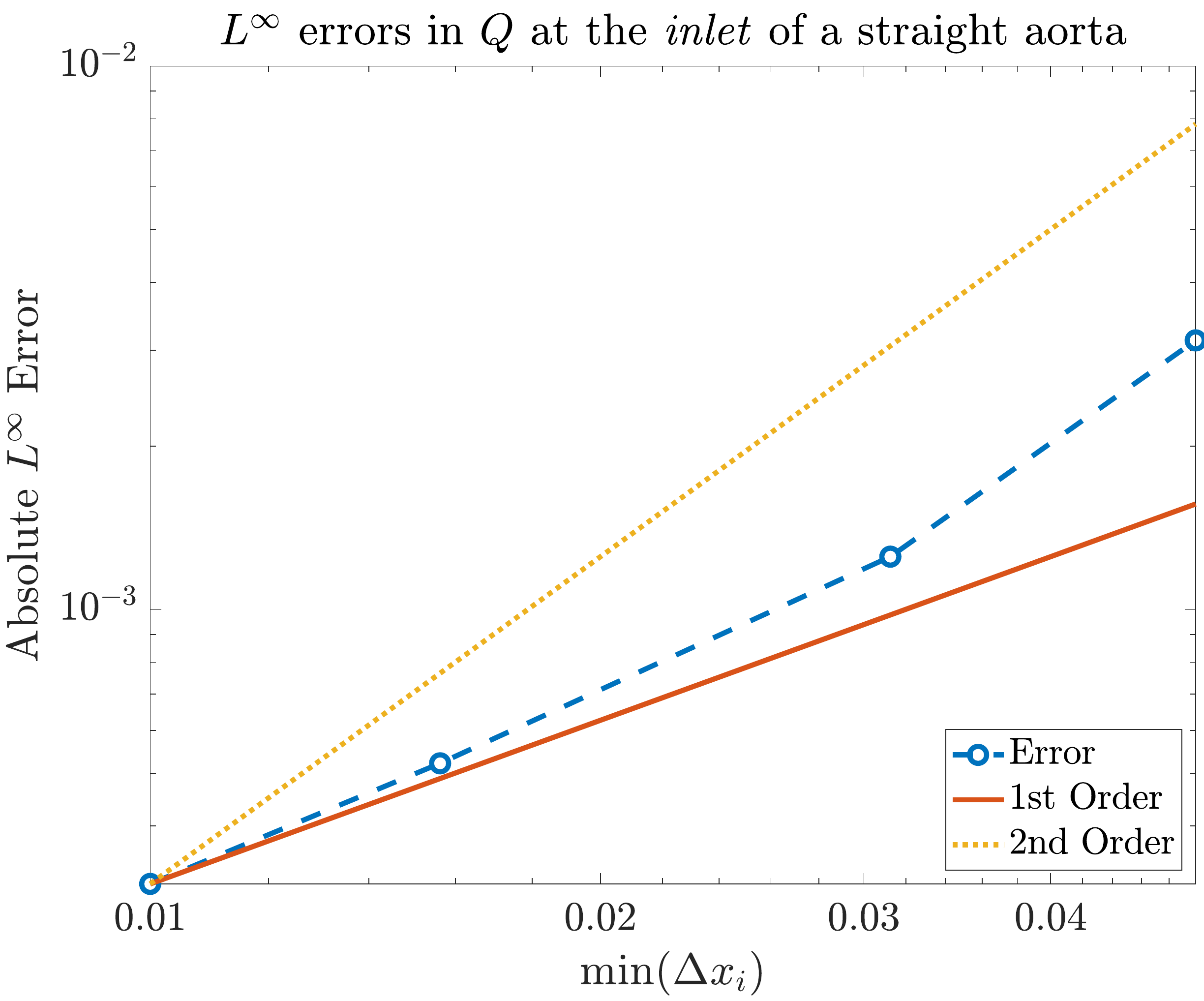}
\caption{\label{fig:qin} (Left) Flow profiles at the inlet for successively-refined discretizations of a straight aorta as produced by the LV-elastance model. (Right) The corresponding $L^\infty$ errors (relative to the finest solution).}
\end{figure}

\begin{figure}
\centering
\includegraphics[width=.485\textwidth,valign=m]{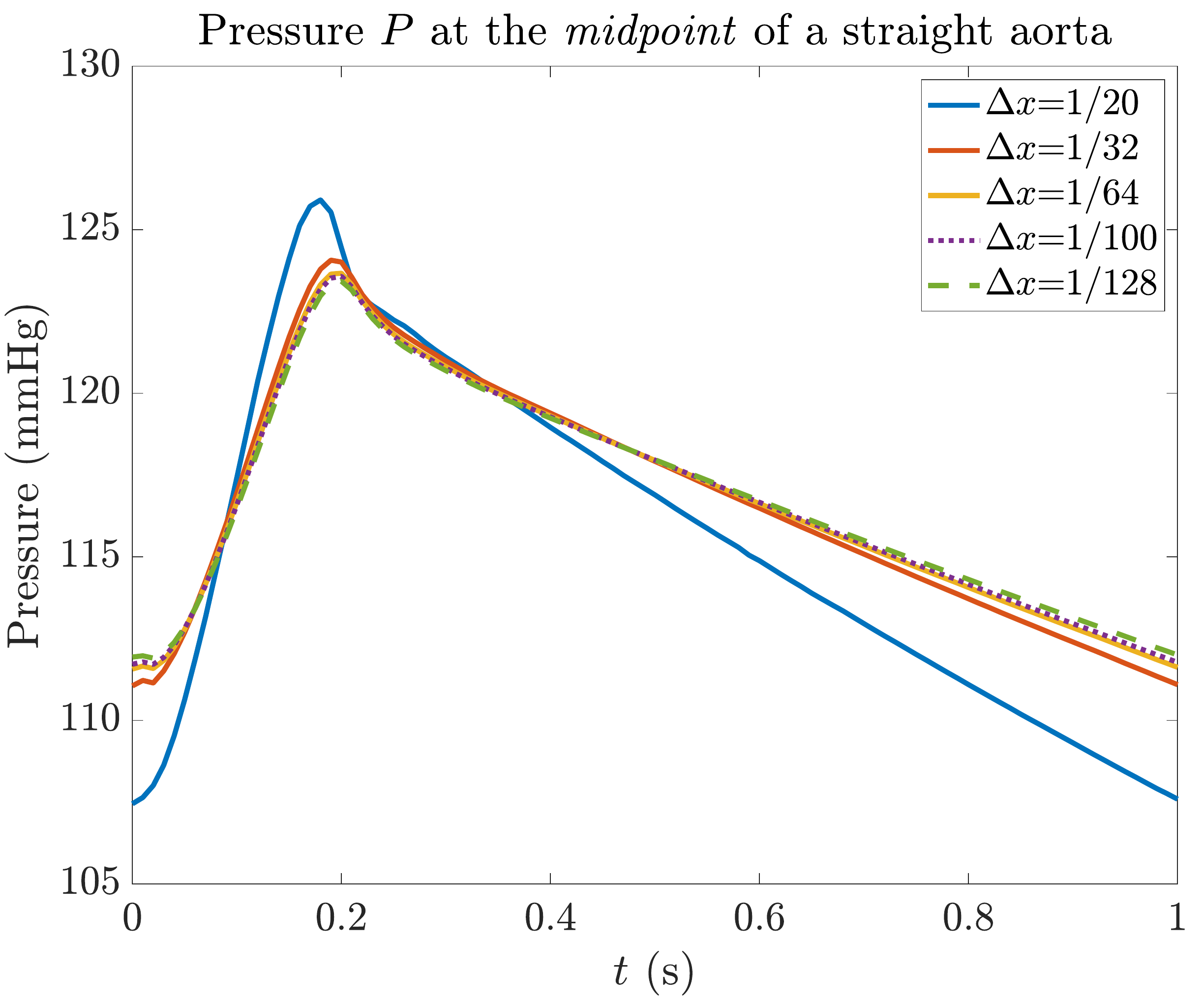}\hfill\includegraphics[width=.485\textwidth,valign=m]{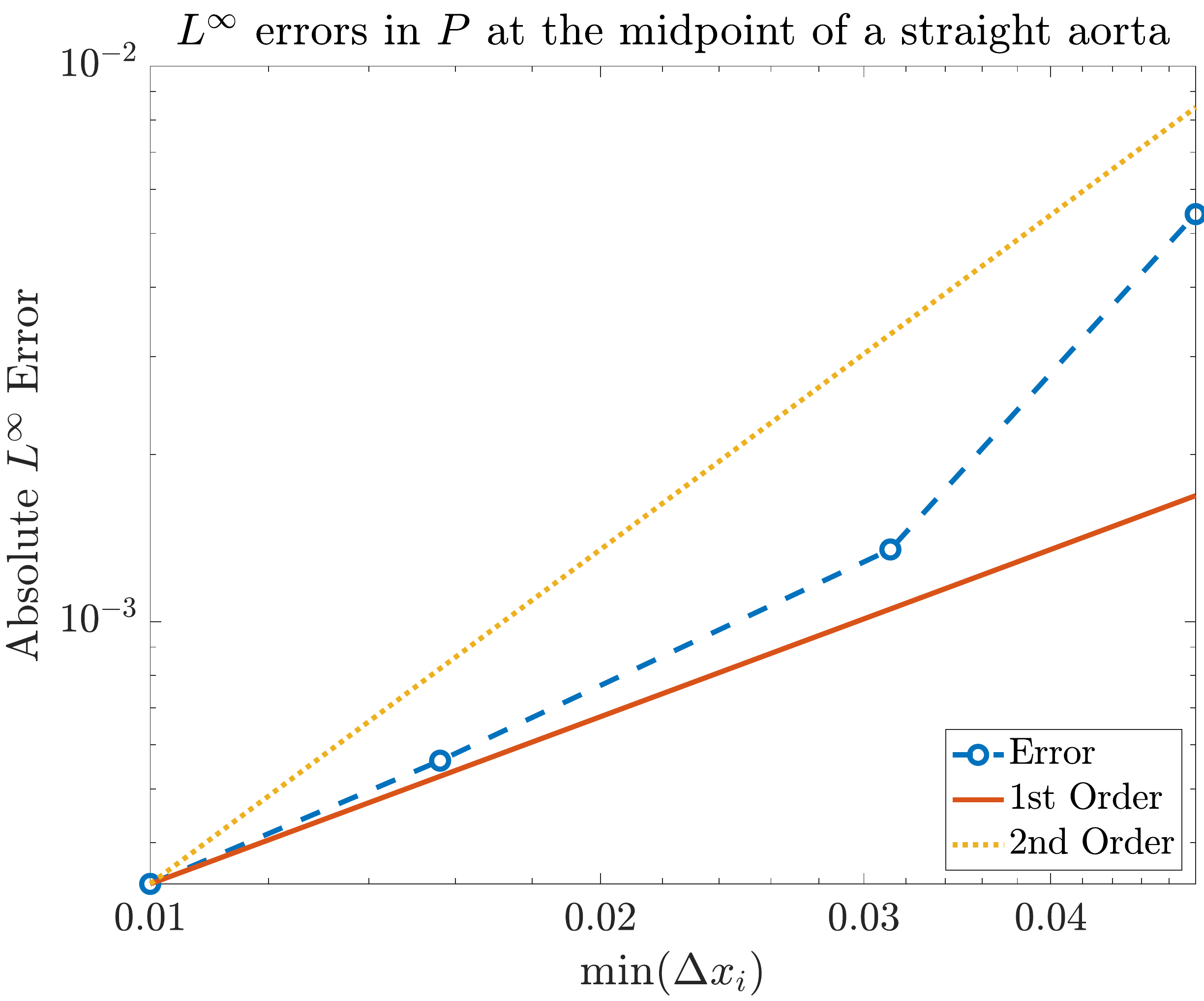}
\caption{\label{fig:pmid} (Left) Pressure profiles at the midpoint for successively-refined discretizations of a straight aorta as produced by the LV-elastance model. (Right) The corresponding $L^\infty$ errors (relative to the finest solution).}
\end{figure}

\subsection{An example physiological case: wall shear stress in the aorta}\label{sec:casestudy}

A predominant effect of advanced congestive heart failure (CHF) is reduced blood flow in the aorta that results from a reduction in cardiac output (CO) and a low ejection fraction (in almost half of the patients). Many factors can influence the heart's pumping ability, including those related to the direct coupling between the LV and the arterial system~\cite{berger,berger2,campbell}. The hybrid ODE-Dirichlet boundary condition considered in this paper has been chosen for its ability to model the non-stationary and nonlinear effects of such complex coupling (which is expressed as an alternating boundary condition between systole---an ODE---and diastole---a Dirichlet condition---as described in Section~\ref{sec:governingLV}). The 0D elastance- (compliance-)based LV model enables the generation of physiological pressure and flow waveforms that can account for different contractilities and cardiac outputs, and its corresponding coupling to 3D lattice Boltzmann equations can enable investigation of the heart's influence on corresponding 3D fluid-structure effects such as those related to near-wall shear stress (WSS). Indeed, mechanical experiments~\cite{gharib2003correlation} and both in-vivo/in-vitro studies~\cite{moore1994fluid,pedersen1993two,oyre1997vivo, pedersen1999quantitative} have shown there can exist negative WSS corresponding to a retrograde flow during a substantial interval within a cardiac cycle, and such effects have been strongly correlated with the state of CHF~\cite{gharib2003correlation}.

As a demonstration of the applicability of the proposed solver towards exploring these parameters for studying pathophysiological conditions in the cardiovascular system (e.g., CHF), a computational model of a simplified 3D aorta that includes carotid and renal branches is considered and illustrated in Figure~\ref{fig:3DModel} (left), where the elastic wall is discretized by finite elements and the fluid by lattice Boltzmann as described in Section~\ref{sec:IBalg}. For an effective aortic diameter $D$ (taken to be unity in the non-dimensionalized configuration), such a domain corresponds to Cartesian coordinates given by $\bm{x}\in[0, 18D]\times[-1.5D,1.5D]\times[0,8D]$ for a non-dimensionalized $D=1$ (corresponding to $24$ mm).  For the compliant wall, a linear elastic material with Young's modulus of $E=0.5$ MPa and a wall thickness corresponding to $h=1$ mm is considered. The complete fluid-structure (immersed boundary) solver, where a no-slip condition is imposed at the fluid-structure interface, is coupled to the 0D LV-elastance heart model (Figure~\ref{fig:2}) at the inlet and a Windkessel ODE at the outlet (see Section~\ref{sec:governingLV}). At all peripheral branch outlets, extension tube boundary models~\cite{pahlevan2011physiologically} are employed. Figure~\ref{fig:3DModel} (right) presents the corresponding normalized velocity magnitudes produced by a simulation that employs discretizations of $\Delta x = 1/32 D, \Delta t = 1/50000$ s and is advanced up to a time $T = 5$ s (where $1$ s corresponds to the period of a cardiac cycle). The LV parameters (including the compliance function) and the Windkessel lumped parameters are adopted from previous work~\cite{amlanipahlevan} and correspond to a healthy case with normal contractility. Additionally, Figure~\ref{fig:3DPin} illustrates the expected physiological characteristics of the LV and aortic pressures, particularly the equality during systole (in the absence of a diseased valve condition) between ventricular pressure $P_v(t)$ and the aortic pressure $P(\bm{x},t)$ at the coupled boundary. Figure~\ref{fig:3DP_all} (left) further demonstrates that the simulations capture the expected increase in pressure amplitude as the LV-sourced waves propagate downstream.   Figure~\ref{fig:3DP_all} (right) provides the corresponding flow profiles as simulated at the inlet, midpoint and outlet of the 3D aorta, demonstrating the physiologically-expected decrease in amplitude as flow propagates downstream.

\begin{figure}
\centering
\includegraphics[width=.485\textwidth,valign=m]{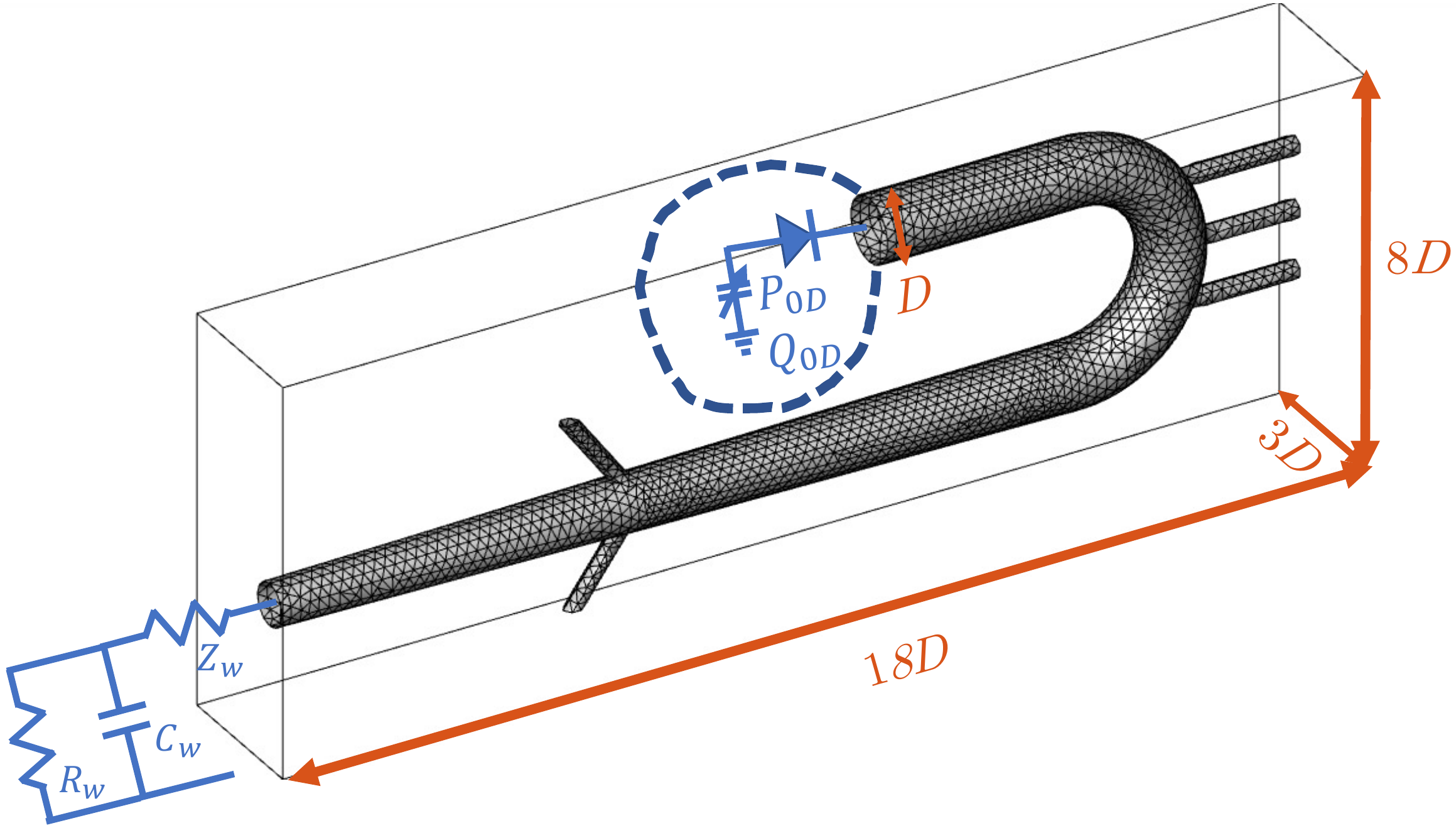}\hfill \includegraphics[width=.485\textwidth,valign=m]{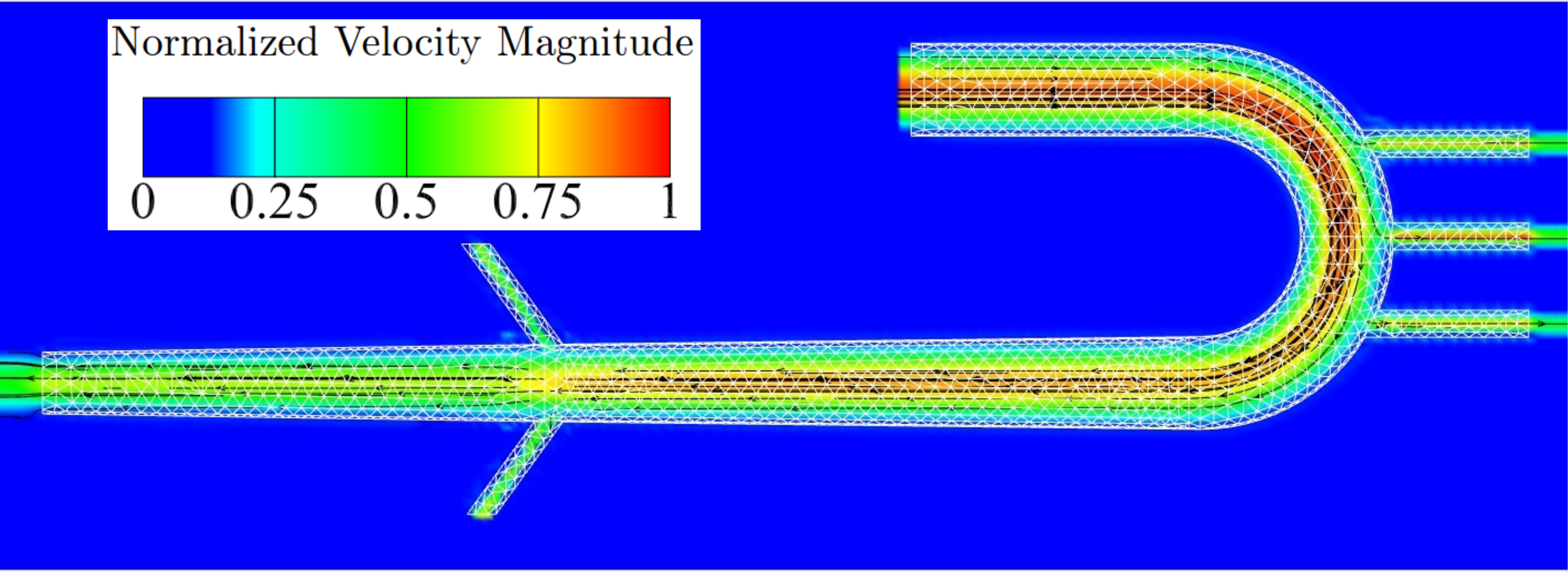}% Here is how to import EPS art
\caption{ (Left) Diagram of a physiologically-relevant 3D aortic domain (with carotid and renal branches) coupled to an LV-elastance model at the aortic inlet and a lumped-parameter Windkessel model at the aortic outlet. (Right) A temporal snapshot of the normalized flow velocity magnitude produced by the solver.\label{fig:3DModel}}
\end{figure}

\begin{figure}
\centering
\includegraphics[width=.485\textwidth,valign=m]{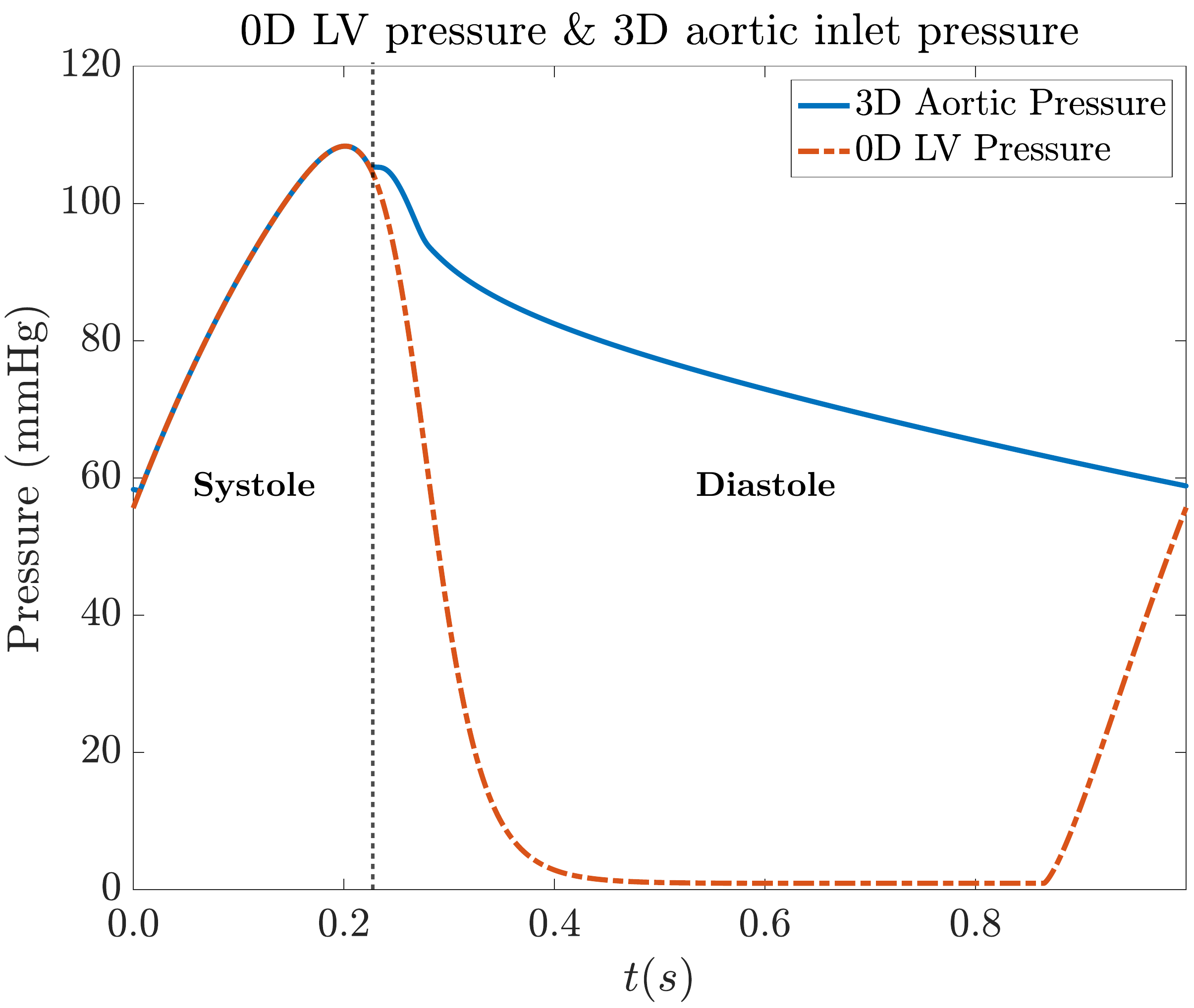}
\caption{\label{fig:3DPin} Aortic pressure at the inlet (blue) and the corresponding ventricular pressure (dashed red) for healthy patient parameters, simulated by the 3D FSI solver. As expected, aortic inlet pressure is equal to LV pressure during the systolic phase (when the valve is opened).}
\end{figure}

\begin{figure}
\centering
 \includegraphics[width=.485\textwidth,valign=m]{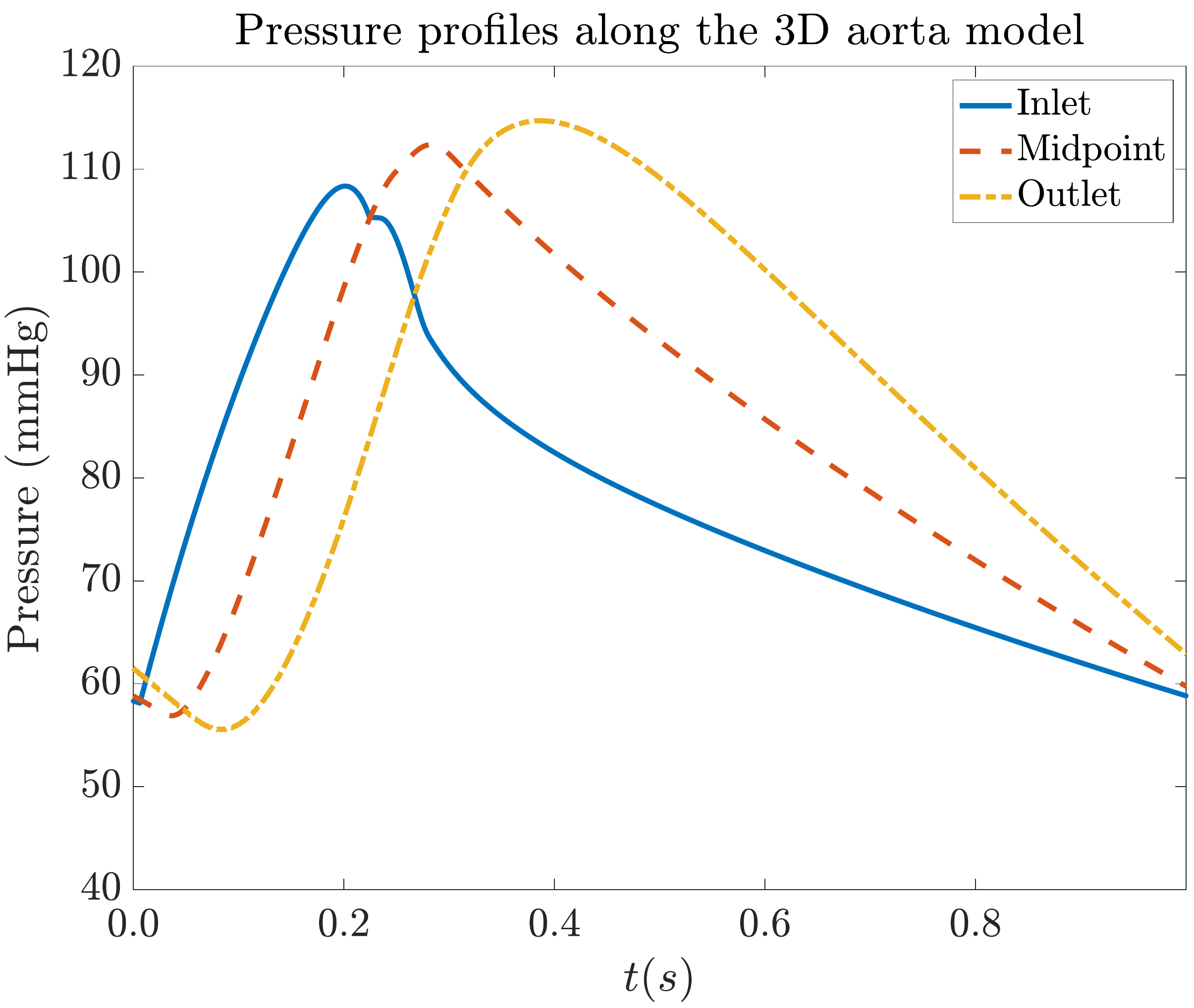}\hfill% Here is how to import EPS art
 \includegraphics[width=.485\textwidth,valign=m]{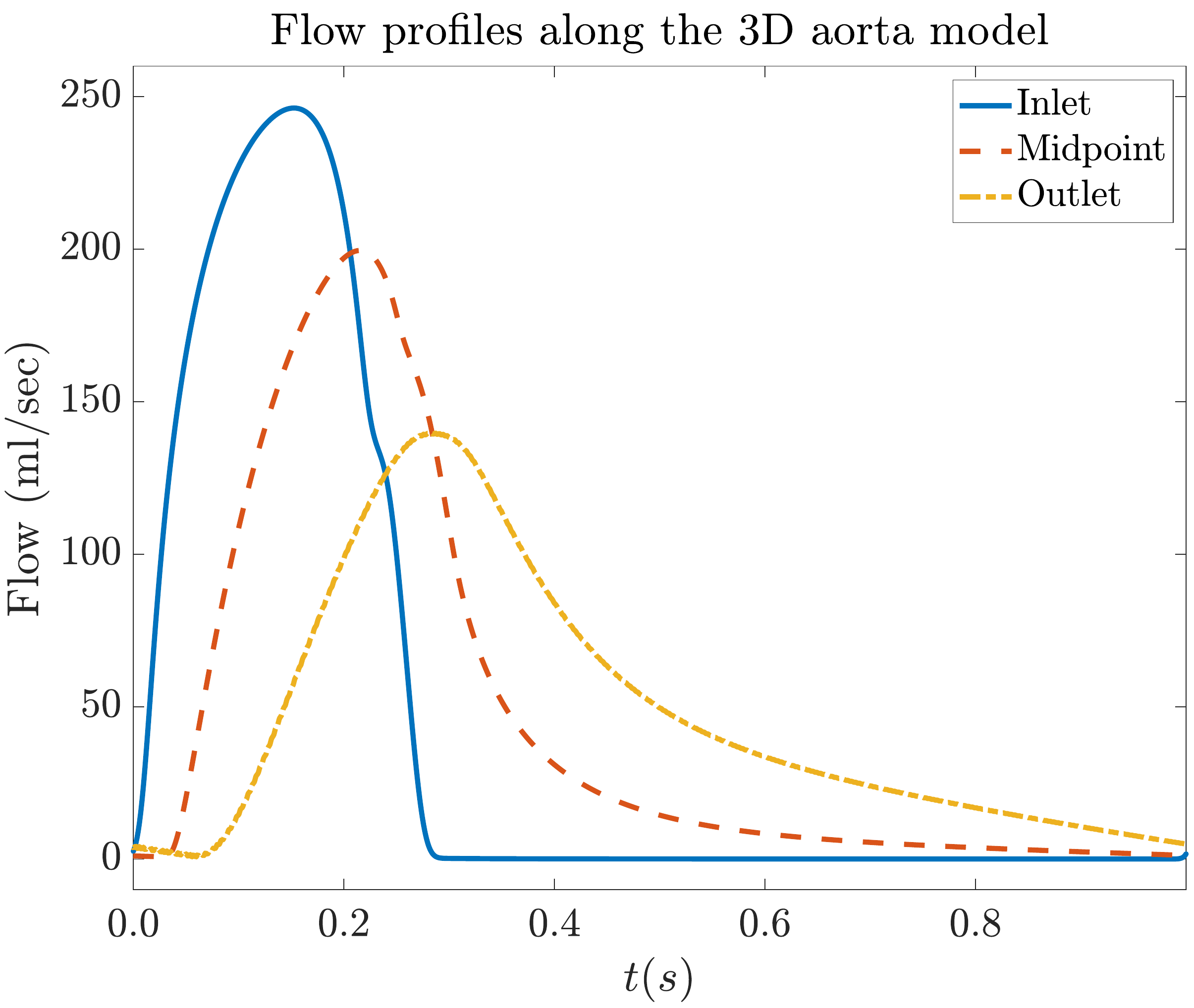}% Here is how to import EPS ar
\caption{\label{fig:3DP_all} (Left) Pressure profiles at the inlet, midpoint and outlet of the 3D aorta model, demonstrating the expected amplification as flow propagates downstream. (Right) Corresponding flow profiles simulated at the inlet, midpoint and outlet, demonstrating the expected decrease in flow amplitude as the wave propagates downstream.}
\end{figure}

\begin{figure}
\centering

\includegraphics[width=.485\textwidth,valign=m]{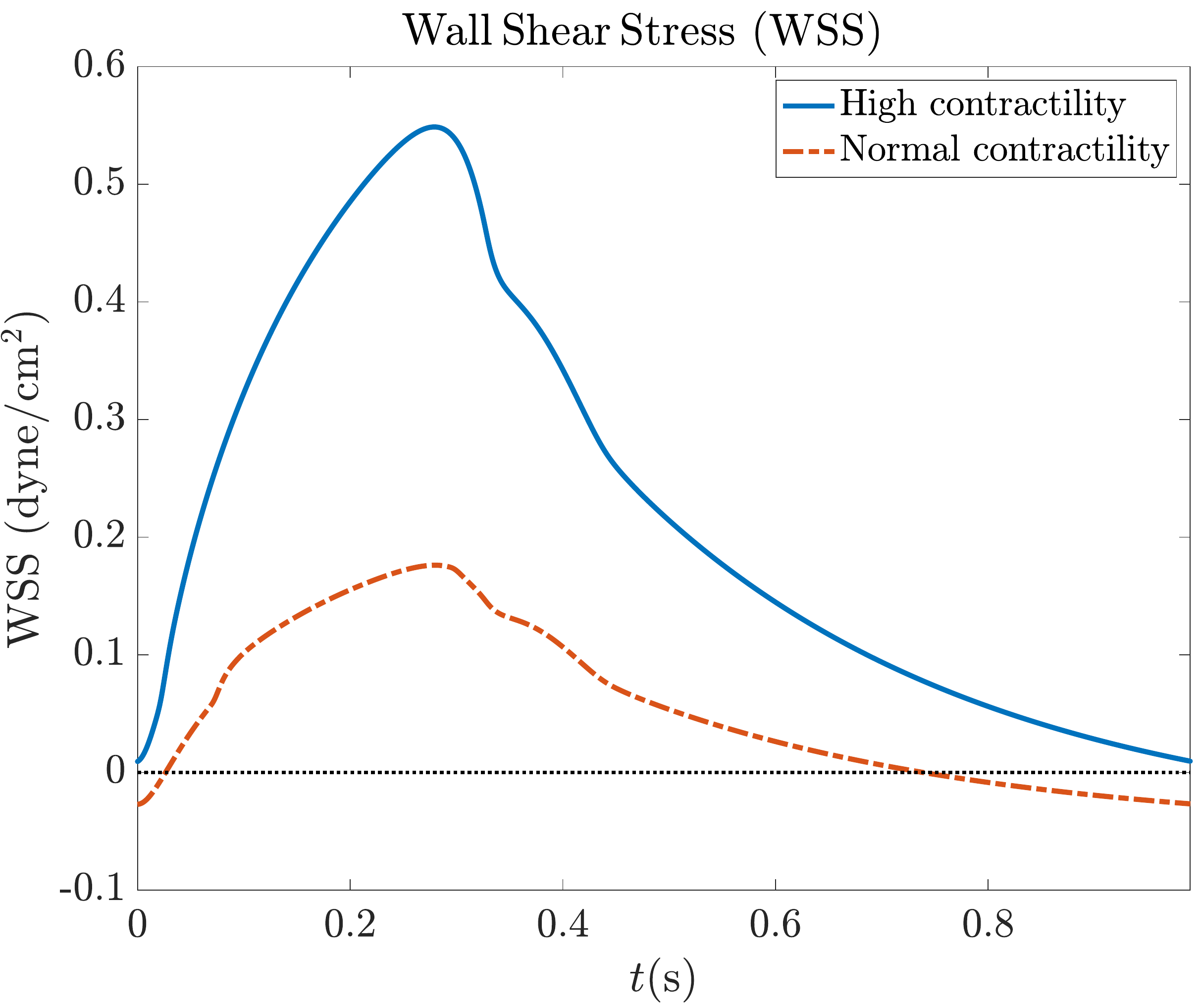}% Here is how to import EPS art
\caption{\label{fig:3DWSS} Simulated wall shear stress (WSS), for both normal and high contractility cases, at a location between the midpoint and outlet of the 3D aorta model.}
\end{figure}

In investigating WSS (as a relevant hemodynamic biomarker in CHF~\cite{gharib2003correlation}), two contractility cases (representing a low flow rate and a high flow rate) can be considered employing the same Womersely number $\text{W}_\text{o} = 16$. For normal contractility, the end-systolic LV elastance $E_{es}$~\cite{amlanipahlevan} is set to $1.76$ mmHg/ml, which corresponds to a cardiac output of $\text{CO}\approx 3.5$ l/min and is in accordance with values employed in experimental studies~\cite{gharib2003correlation}. For the high contractility scenario, $E_{es}$= $2.75$ mmHg/ml. Again, for both cases, the Womersely number $\text{W}_\text{o} = 16$ (corresponding to a heart rate of $\text{HR}=60$ BPM~\cite{gharib2003correlation}) is fixed.  Figure~\ref{fig:3DWSS} presents the corresponding wall shear stress (WSS) for both normal and high contractility cases, as calculated through the fluid points next to the solid wall~\cite{gharib2003correlation} via the expression
\begin{equation}
  \text{WSS}(\bm{x},t) = \mu \frac{\partial v_1}{\partial x_3}(\bm{x},t),
\end{equation}
where $\mu$ is the fluid viscosity (corresponding to 3.5 centipoise), $v_1$ is the simulated axial velocity, and $x_3$ is the dimension normal to the wall. As expected~\cite{gharib2003correlation}, a negative WSS (corresponding to a retrograde flow) is evident for the normal contractility case parameters, and such effects dissappear in the high contractility case since the corresponding flow rate is very high. These results are in agreement with the experimental results presented in Gharib and Beizaie~\cite{gharib2003correlation}.

\section{{Conclusion}}\label{sec:conclusions}

This work presents a direct 0D-3D coupling for dynamic (ODE-based) boundary conditions applied to lattice Boltzmann solvers for hemodynamic flow. Benchmark performance studies and a physiological case of wall shear stress in a simplified 3D aorta are treated in order to validate the proposed methodology and its implementation. In particular, this work treats a most complicated configuration of such coupling conditions: a hybrid non-stationary ODE-Dirichlet boundary condition. Such a methodology produces a physiologically-accurate hemodynamics solver (with a heart model) for studying wave propagation and pulsatile blood flow in arterial vessels. The methodology introduced in this paper can be easily extended to non-switching ODE conditions such as 0D lumped parameter models (e.g., Windkessels for truncating vasculature at vessel outlets), as well as to other methods for treating fluid-structure interactions (facilitated here by an immersed boundary method). Such a direct 0D-3D coupling with the proposed regularization of Section~\ref{sec:coupling} can also be applied to any other fluid problem that is governed by lattice Boltzmann equations and that requires direct time-dependent ODEs as boundary conditions.

% \nocite{*}% Show all bib entries - both cited and uncited; comment this line to view only cited bib entries;
% \section{Ethical Statement}
%
% None.
\section{{References}}

\bibliographystyle{elsarticle-num}
\bibliography{references}

\end{document}